\documentclass[aps,prb,twocolumn,nofootinbib,superscriptaddress]{revtex4-2}
\usepackage{graphicx}
\usepackage{amsmath,amssymb}
\usepackage[utf8]{inputenc}
\usepackage[colorlinks = true,
linkcolor = blue,
urlcolor  = blue,
citecolor = blue,
anchorcolor = blue]{hyperref}
\usepackage{braket}
\usepackage[mathscr]{euscript}
\usepackage{mathdots}
\usepackage{color}

\begin{document}
\title{Floquet Non-Bloch Formalism for a Non-Hermitian Ladder: From Theoretical Framework to Topolectrical Circuits}

\author{Koustav Roy$^@$}
\email{koustav.roy@iitg.ac.in}
\affiliation{Department of Physics, Indian Institute of Technology Guwahati-Guwahati, 781039 Assam, India}
\author{Dipendu Halder$^@$}
\email{h.dipendu@iitg.ac.in}
\affiliation{Department of Physics, Indian Institute of Technology Guwahati-Guwahati, 781039 Assam, India}
\author{Koustabh Gogoi}
\affiliation{Department of Physics, Indian Institute of Technology Gandhinagar-Gandhinagar, 382055, Gujarat, India}
\author{B. Tanatar}
\email[Corresponding author: ]{tanatar@fen.bilkent.edu.tr}
\affiliation{Department of Physics, Bilkent University, 06800 Bilkent, Ankara, Turkey}
\author{Saurabh Basu}
\affiliation{Department of Physics, Indian Institute of Technology Guwahati-Guwahati, 781039 Assam, India}

\begin{abstract}
\noindent Periodically driven systems intertwined with non‑Hermiticity opens a rich arena for topological phases that transcend conventional Hermitian limits.
The physical significance of these phases hinges on obtaining the topological invariants that restore the bulk‑boundary correspondence, a task well explored for static non-Hermitian (NH) systems, while it remains elusive for the driven scenario.
Here, we address this problem by constructing a generalized Floquet non‑Bloch framework that analytically captures the spectral and topological properties of time-periodic NH systems.
Employing a high-frequency Magnus expansion, we analytically derive an effective Floquet Hamiltonian and formulate the generalized Brillouin zone for a periodically driven quasi-one-dimensional system, namely, the Creutz ladder with a staggered complex potential.
Our study demonstrates that the skin effect remains robust (despite the absence of non-reciprocal hopping) across a broad range of driving parameters, and is notably amplified in the low-frequency regime due to emergent longer-range couplings.
We further employ a symmetric time frame approach that generates chiral-partner Hamiltonians, whose invariants, when appropriately combined, account for the full edge-state structure.
To substantiate the theoretical framework, we propose a topolectrical circuit (TEC) that serves as a viable experimental setting.
Apart from capturing the skin modes, the proposed TEC design faithfully reproduces the presence of distinct Floquet edge states, as revealed through the voltage and impedance profiles, respectively.
Thus, our work not only offers a theoretical framework for exploring NH-driven systems, but also provides an experimentally feasible TEC architecture for realizing these phenomena stated above in a laboratory.
\end{abstract}

\maketitle
\def\thefootnote{@}\footnotetext{These authors contributed equally to this work}
\section{\label{s1}Introduction}

The interplay between topology and non-Hermiticity \cite{PhysRevLett.116.133903, PhysRevLett.120.146402, Ghatak_2019, PhysRevX.8.031079, PhysRevX.9.041015, Ashida2020, RevModPhys.93.015005} has recently emerged as a captivating frontier in condensed matter physics, unveiling a plethora of intriguing physical phenomena that challenge traditional paradigms.
One such striking manifestation is the emergence of exceptional points \cite{PhysRevB.97.121401, Heiss_2012, PhysRevLett.118.040401} where the Hamiltonian becomes defective and multiple eigenvectors coalesce, giving rise to spectral singularities.
Another significant hallmark of non-Hermitian (NH) systems is the non-Hermitian skin effect (NHSE) \cite{PhysRevLett.121.086803, PhysRevLett.121.026808, PhysRevB.99.201103, PhysRevLett.124.056802, PhysRevLett.124.086801, Halder_2024, Manna2023}, wherein the system becomes extremely sensitive to boundary conditions.
Nevertheless, in order to address the breakdown of Bulk-boundary correspondence (BBC) caused by NHSE, various strategies have been proposed, such as the biorthogonal formulation of eigenstates \cite{Kunst2018}, singular value decomposition approaches \cite{Herviou2019}, and gauge transformations \cite{Ge2019}.
A particularly significant advancement in this regard is the development of the non-Bloch band theory \cite{PhysRevLett.123.066404,Zhang2020,Wu2020}, which generalizes the conventional Brillouin zone (BZ) to a complex-valued generalized Brillouin zone (GBZ).
This formalism enables the precise definition of non-Bloch topological invariants, thereby accurately characterizing non-trivial edge states in NH systems.
These features have sparked intense theoretical and experimental investigations across diverse platforms, including ultracold atoms \cite{El-Ganainy2018, Eichelkraut2013}, electronic circuits \cite{PhysRevLett.114.173902, Helbig2020, Lee2018, PhysRevB.109.115407}, mechanical \cite{Wang}, and acoustic systems \cite{Fleury2015, PhysRevApplied.16.057001}, underscoring the versatility of NH physics as a robust platform for exploring the nexus between topology and non-Hermiticity.

While the exploration of NH topological systems has already unveiled a multitude of exotic physical phenomena, a further dimension of control and novelty can be achieved by extending these studies beyond equilibrium scenarios.
In this context, Floquet engineering \cite{Grifoni1998, Cayssol2013, GomezLeon2013, Goldman2014, Restrepo2016, Rudner2013, Rudner2020} has emerged as a powerful quantum control technique, offering an elegant and systematic framework to tailor and manipulate topological properties in out-of-equilibrium scenarios.
By periodically modulating system parameters, Floquet theory reveals that additional non-trivial states of matter can be realized, which have no static analogue.
Moreover, Floquet engineering has uncovered remarkable features such as the realization of Floquet topological phases with multiple dynamically generated edge states \cite{Agrawal2022, Li2020, Roy2023, Yang2022, Roy2024}, unconventional transport phenomena \cite{SanJose2013prime, Peng2021prime, Liu2019prime, Roy2025}, Floquet Anderson phases featuring localized bulk and protected edge modes \cite{Roy2024prime}, and even discrete time crystalline phases that spontaneously break the time-translation symmetry \cite{Pan2020, Wang2022}.
These discoveries underscore the transformative potential of periodic driving in broadening the landscape of accessible quantum phases and controlling exotic phenomena unattainable within a static framework.

Building upon the profound capabilities of Floquet engineering to realize exotic topological phases, it becomes imperative to investigate how these driving-induced phenomena manifest in specific lattice models, especially when combined with NH effects \cite{PhysRevB.108.L220301, PhysRevLett.132.063804, Ji_2024,Ji2025}.
Among various TB models, the Creutz ladder, being a quasi-one-dimensional (1D) system \cite{Creutz1999, Gholizadeh2018}, emerges as an exceptionally powerful and versatile model due to its direct experimental realizability in cold atomic systems \cite{Kang2020, Li2013}.
Originally introduced to study chiral fermions, the Creutz ladder consists of two rungs of lattice sites connected via diagonal, vertical, and horizontal hoppings, encapsulating effective two-dimensional topological aspects and unique symmetry classifications \cite{Hugel2014}.
Furthermore, the localization of zero-energy modes in the Creutz ladder is influenced by Aharonov-Bohm caging, wherein the destructive interference confines particles within a finite region.
This dual protection ensures the robustness of edge modes against perturbations.
Recent studies have proposed various modifications of the Creutz ladder to uncover rich information in the realms of many-body interactions \cite{Junemann2017} and drive-induced localization phenomena \cite{DiLiberto2019, Kuno2020, Roy2023}.
Despite extensive studies on the Creutz ladder in both static and periodically driven scenarios \cite{LiangLi2022, Zhou2020}, an explicit investigation into NHSE and the associated GBZ framework remains largely unexplored.
Our primary aim is to unravel the interplay between non-Hermiticity and periodic driving, and capture it through the Floquet GBZ formalism.

Moreover, to bridge the gap between theoretical predictions and experimental implementation, we further seek to translate our driven model into a realizable physical system. Among various experimental platforms, topolectrical circuits (TECs) have evolved as a particularly powerful and versatile tool in experiments, relying on their ability to map tight-binding (TB) Hamiltonians onto circuit Laplacians \cite{PhysRevLett.114.173902, Lee2018, PhysRevB.99.161114, PhysRevResearch.3.023056, 7t7k-qg49, YANG20241, 10.1063/5.0265293}.
The zero-energy edge (skin) modes can be realized through the impedance (voltage) profile via exciting specific nodes within the circuit network.
By adjusting electrical components and connection configurations, TECs offer remarkable flexibility to engineer and explore a wide range of topological characteristics, enabling precise control over the system parameters.
This unique flexibility, combined with their direct measurement capabilities, makes TECs an outstanding platform to realize theoretical models, particularly those emerging in NH systems.
Notably, while static Hamiltonians have been widely explored, it remains a crucial open challenge to realize time-dependent perturbations, such as periodically driven systems, within the framework of TECs.
Thus, we intend to systematically characterize the topological phases of the NH Creutz ladder using the GBZ, construct a comprehensive phase diagram as a function of different driving and NH parameters, and finally validate our theoretical findings by designing a TEC.

The remainder of this paper is organized as follows. Sec.~\ref{sec:level2} presents an overview of the static NH Creutz ladder and its key topological features, followed by which we introduce the Floquet formalism and construct the effective time-independent Floquet Hamiltonian for the driven model.
Sec.~\ref{sec:level3} explores the NH characteristics of the driven system and utilizes the Floquet GBZ to restore BBC.
In Sec.~\ref{sec:level4}, we propose an experimental realization of our results using TECs.
Finally, Sec.~\ref{sec:level6} concludes with a summary of our findings and future outlook.
\begin{figure}[t]
         \includegraphics[width=\columnwidth]{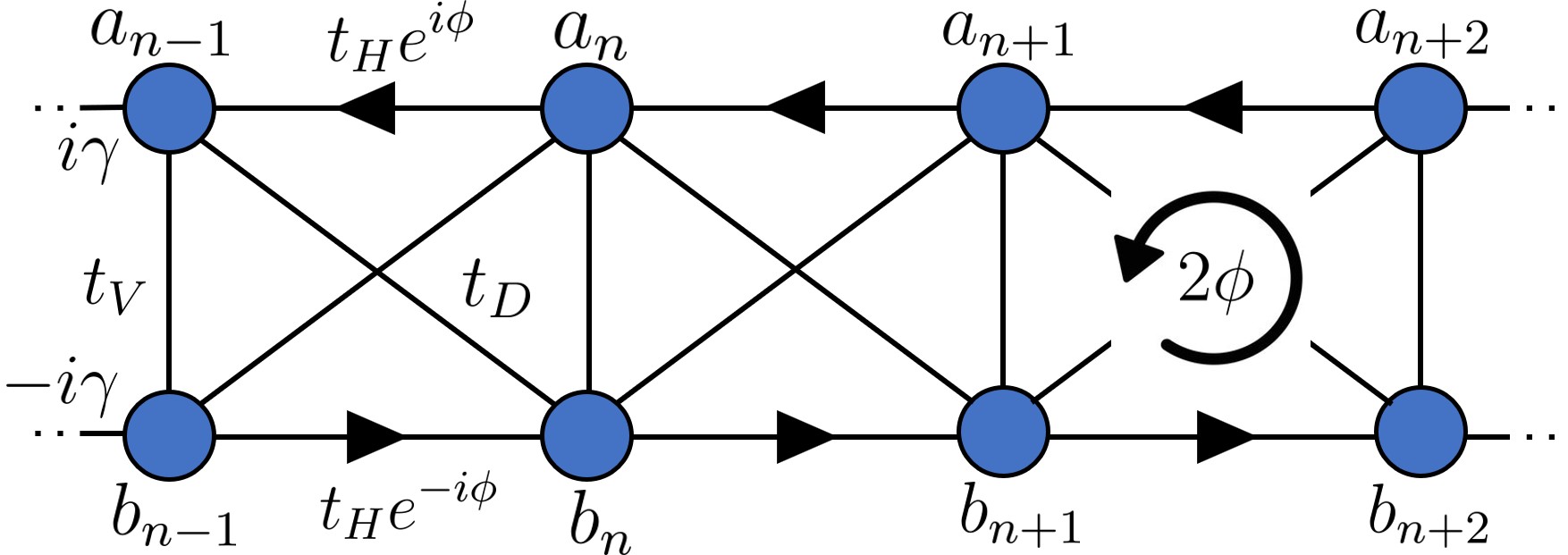}
\caption{Schematic representation of the quasi-1D Creutz ladder, where $a_n$ and $b_n$ denote the two distinct sublattices. The different hopping amplitudes, $t_H$, $t_V$, $t_D$, denote the horizontal, vertical, and diagonal hoppings, respectively.}
\label{fig:1}
\end{figure}
\section{\label{sec:level2}Driven NH Creutz ladder: Floquet Topology and skin effect}

The Creutz ladder is composed of two parallel chains (or `rungs') of lattice sites, interconnected via diagonal $(t_D)$, vertical $(t_V)$, and horizontal $(t_H)$ hopping amplitudes, as illustrated in Fig.~\ref{fig:1}.
A distinctive feature of this model is the presence of a magnetic flux threading the ladder, thereby introducing an additional degree of freedom through the Peierls phase ($\phi$) associated with horizontal hopping.
Each unit cell contains two sublattice sites, denoted by $a_n$ and $b_n$. The corresponding Hamiltonian in momentum space takes the form \cite{Creutz1999, Gholizadeh2018},
\begin{align}
H(k)=2t_{H}[\cos{k}\cos{\phi}\,\sigma_0&+\sin{k}\sin{\phi}\,\sigma_z]\nonumber\\&+(t_V+2t_{D}\cos{k})\,\sigma_x.
\label{eq:2}
\end{align}
Here, the operators $\sigma_i = x, y, z$ correspond to the Pauli matrices, and $\sigma_0$ represents $2 \times 2$ identity matrix. Further, a magnetic flux $\phi$ threads each plaquette of the ladder, which is related to the hopping phase via $2\phi = \frac{\Phi}{\Phi_0}$, with $\Phi_0$ denoting the magnetic flux quantum. Interestingly, at the special points $\phi = \pm \pi/2$, the Creutz ladder becomes mathematically equivalent to the Su-Schrieffer-Heeger (SSH) model \cite{SuSchriefferHeeger1979}, provided a rotation of the spin basis is performed via the transformation, $\sigma_y \rightarrow \sigma_z \rightarrow -\sigma_y.$

At this stage, it is essential to highlight the symmetries inherent in the model \cite{PhysRevB.92.085118, PhysRevB.83.245132, PhysRevX.7.031057}.
The system exhibits inversion symmetry concerning a horizontal axis centered between the two legs of the ladder, captured by the condition $\sigma_x H(k)\sigma_x = H(-k)$.
In addition, the model supports a chiral symmetry at the special values $\phi = \pm \frac{\pi}{2}$, expressed as $\sigma_y H(k)\sigma_y = -H(k)$.
Remarkably, despite the presence of a magnetic flux, the system retains an effective time-reversal symmetry (provided $\phi = \frac{\pi}{2}$), given by $\mathcal{K} H(k)\mathcal{K}^{-1} = H(-k)$, with $\mathcal{K}$ being the complex conjugation operator.
Finally, for $\phi = \frac{\pi}{2}$, the model also respects a particle-hole symmetry, described by $\sigma_z H(k)\sigma_z = -H(-k)$.

Unlike the standard SSH chain, the Creutz ladder features a broader set of tunable parameters, opening the door to a richer landscape of phenomena when NH effects are introduced.
In particular, we examine the impact of adding a sublattice-dependent imaginary onsite potential, represented by the term
\begin{equation}
H_{\gamma} = \sum_{n} i (\gamma_A\;a_n^\dagger a_n + \gamma_B\;b_n^\dagger b_n).
\label{eq:Ham3}
\end{equation}
In general, such an NH potential $(i\gamma)$ does not induce non-reciprocity and, therefore, does not give rise to the NHSE.
However, when a staggered configuration is imposed, specifically, $\gamma_A = -\gamma_B = \gamma$, and the pseudospin basis is rotated as $\sigma_y \rightarrow \sigma_z \rightarrow -\sigma_y$., an effective asymmetry emerges in the intra-cell hopping terms involving $t_V$ and $\gamma$.
This asymmetry mimics that found in non-reciprocal extensions of the SSH model \cite{PhysRevLett.121.086803, Halder_2023}, thereby enabling the onset of NHSE in the Creutz ladder.
In the following sections, we validate this mathematical framework using TECs and demonstrate that an analogous phenomenon to the skin effect can arise without the introduction of any explicit non-reciprocity in the TEC.
With the introduction of $H_{\gamma}$ to the Hermitian Hamiltonian $H$ in Eq.~\eqref{eq:2}, the total Hamiltonian of the system can be written as,
\begin{equation}
    H_{\text{0}}=H +H_{\gamma}
    \label{eq:Ham4}
\end{equation}

At this stage, the addition of periodic driving adds a new layer of complexity to the system, substantially enriching its topological behavior. A powerful framework to analyze such time-dependent systems is the Floquet formalism \cite{Grifoni1998,Cayssol2013,GomezLeon2013,Goldman2014,Restrepo2016,Rudner2013,Rudner2020}, which enables the construction of an effective time-independent Hamiltonian that governs the time evolution of the system at discrete time intervals. 

\begin{figure}[t]
         \includegraphics[width=\columnwidth]{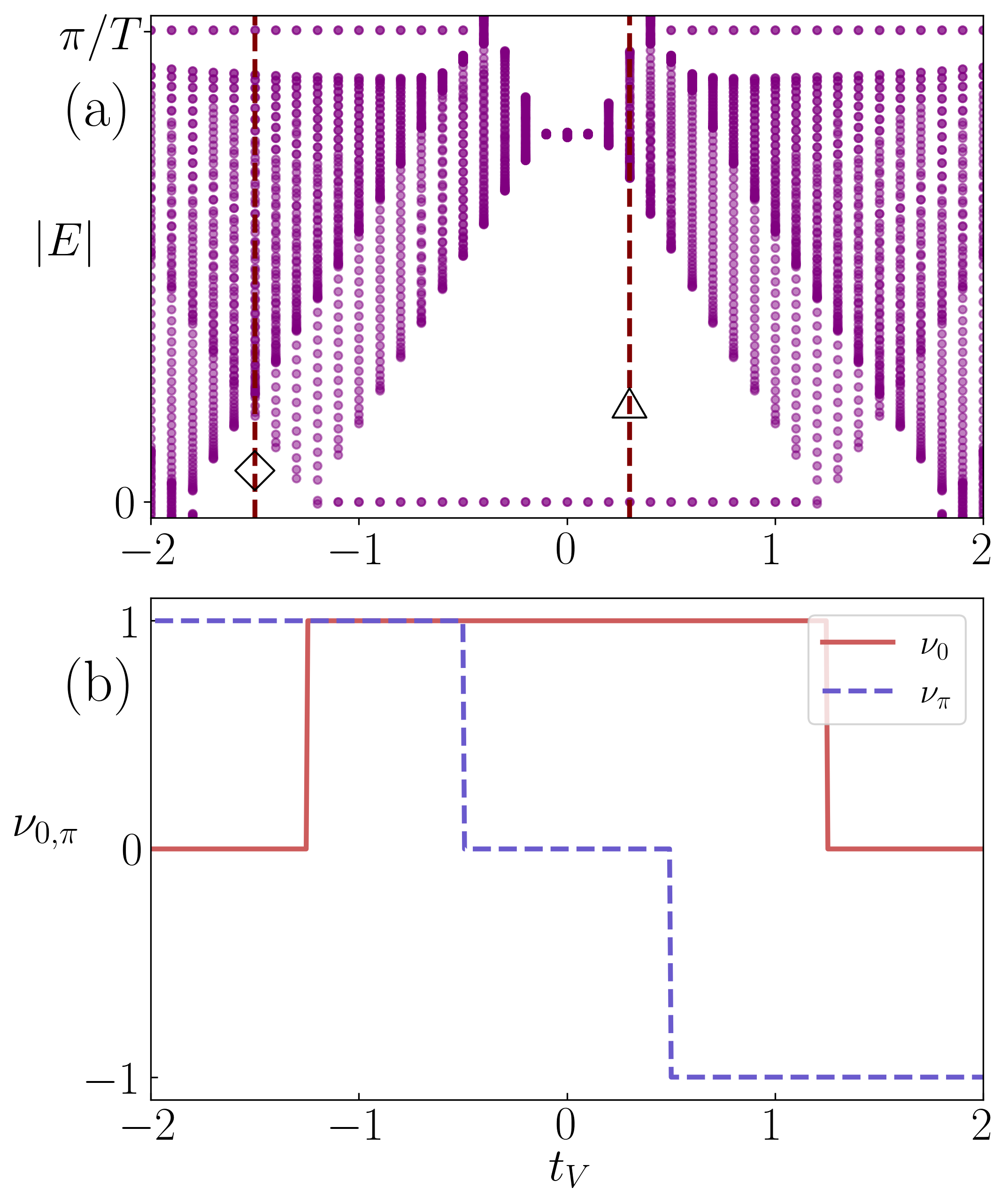}
\caption{Panel (a) shows the Floquet quasi-energy spectrum corresponding to the harmonic drive, plotted as a function of the vertical hopping, $t_V$. Panel (b) demonstrates the bulk invariants $\nu_{0,\pi}$ evaluated from the coordinates of GBZ, which correctly coincides with panel (a). The rest of the parameters are chosen as, $t_H=0.6$, $t_D=0.6$, $V_0=0.3$, $\gamma=0.4$, and $\omega=3$.}
\label{fig:2}
\end{figure}

In the following, we focus on the introduction of a harmonic drive, outlining the associated technical challenges and the strategies required to restore BBC in this periodically driven NH setting. We begin by considering a harmonic modulation of the vertical hopping term, described by
\begin{equation}
t_{V}(t) = (2V_0 \cos{\omega t} + t_V) .
\end{equation}
Here, $V_0$ is the driving strength and $\omega=2\pi/T$ is the driving frequency. While alternative driving protocols such as step drives, implemented via certain periodic quenches, are also possible, their impact on the preservation of BBC has already been systematically analyzed in earlier works \cite{Roy2025Creutz}. However, the harmonic driving case remains relatively unexplored. Moreover, a key motivation here is the experimental feasibility, since one of our ultimate objectives is to design an electrical circuit that can emulate the dynamics of this model. While implementing other driving schemes (which, anyway, require completely different frameworks) in such physical platforms can be technically challenging and less feasible. Harmonic driving, on the other hand, when analyzed in the suitable frequency domain, offers a more tractable framework. It not only simplifies the theoretical treatment of restoring BBC but also facilitates practical realization in TECs
\par 
In order to tackle such time-periodic systems, 
one convenient way is to construct the effective Hamiltonian via working in the frequency domain, which embeds the problem in an extended Hilbert space $\mathscr{R}\otimes\mathscr{T}$, where, $\mathscr{R}$ denotes the usual Hilbert space, while $\mathscr{T}$ is the space of $T$-periodic functions spanned by $\{e^{-im\omega t}\}$.
This yields the following form of the Floquet effective Hamiltonian, $H_F$ as,
\begin{equation}
H_F=\sum_{m,m^\prime} \Big( m\omega \delta_{m,m^{\prime}} + \frac{1}{T} \int_{0}^{T} dt H(t) e^{-i(m-m^\prime)\omega t} \Big).
\end{equation} 
Consequently, in matrix notation the Floquet Hamiltonian can be generally represented as,
\begin{equation}
\label{eq:9}
    \bra{m} H_F \ket{m^{\prime}} = m \omega \delta_{m,m^{\prime}} + E_0 \delta_{m,m^{\prime}} + H_{m-m^{\prime}},
\end{equation}
where $E_0$ is the energy of the static Hamiltonian, and the Fourier elements $H_{\pm m}$ are expressed as the second term in the right-hand side of Eq. \eqref{eq:9}.
The elements $H_{|m|}$, except for $m=0,\pm1$ vanish owing to the mathematical form of the drive.
Moreover, the spectrum of $H_F$ requires the full infinite-dimensional matrix. In practice, however, one can truncate the matrix to a finite number of replicas, whose extent depends on the drive frequency $\omega$. When $\omega$ is larger than the system bandwidth $D$ (defined as $D = 2 (|t_V| + 2|t_D|$), higher-order couplings ($|m|>1$) are strongly suppressed, so that keeping only a few replicas suffices to capture the essential physics. In the following, we use this truncated Floquet Hamiltonian to analyze the quasi-energy spectrum and demonstrate the emergence of the NHSE.
\par Fig.~\ref{fig:2}(a) represents the absolute value of the quasi-energy spectrum obtained by diagonalizing Eq.~\eqref{eq:9} with an appropriate truncation of the off-diagonal Fourier components. The spectrum exhibits the coexistence of both zero and $\pi$ modes as a function of $t_V$, with other parameters fixed at $t_H = t_D = 0.6$, $\omega = 3$, $V_0 = 0.3$, and $\gamma = 0.4$. On the other hand, Fig.~\ref{fig:2}(b) illustrates the behavior of the bulk invariants $\nu_0$ and $\nu_\pi$, which indicate the presence of zero and $\pi$-modes, respectively. These results are in complete agreement with the quasi-energy spectrum shown in Fig.~\ref{fig:2}(a), confirming that we have successfully restored the broken BBC. This has been achieved within a generalized Floquet non-Bloch framework, which requires explicit analytical expressions for the matrix coefficients of the driven Hamiltonian. In the following section, we outline how one can approximate the corresponding $d$-vectors from the driven Hamiltonian. Before doing so, however, we first discuss the role of drive-induced skin modulation by focusing on two representative points highlighted in Fig.~\ref{fig:2}(a), namely $t_V = 0.3$ and $t_V = -1.5$. In both cases, as we see, the system exhibits NHSE (see Fig.~\ref{fig:3}); however, while the wavefunctions are localized at the left edge for $t_V = 0.3$ (Fig.~\ref{fig:3}(a)), they are localized at the right edge for $t_V = -1.5$ (Fig.~\ref{fig:3}(b)). This implies the breakdown of conventional Bloch theory, necessitating the use of non-Bloch band theory to correctly characterize the system. 
\section{\label{sec:level3}Floquet GBZ via Magnus expansion: non Bloch invariants}
In general, constructing the GBZ in a Floquet setup is highly non-trivial, as it requires explicit expressions for the $d$-vectors in the driven scenario \cite{footnote}, that are difficult to extract numerically from the infinite-dimensional Floquet Hamiltonian in Eq.~\eqref{eq:9}. To overcome this challenge, we adopt a rotating frame approximation followed by a high-frequency expansion, also known as the \textit{Magnus expansion} \cite{Benito2014, EckardtAnisimovas2015, Wang2020}, which enables us to derive the expressions for the components of the effective $d$-vectors.
\begin{figure}[t]
         \includegraphics[width=\columnwidth]{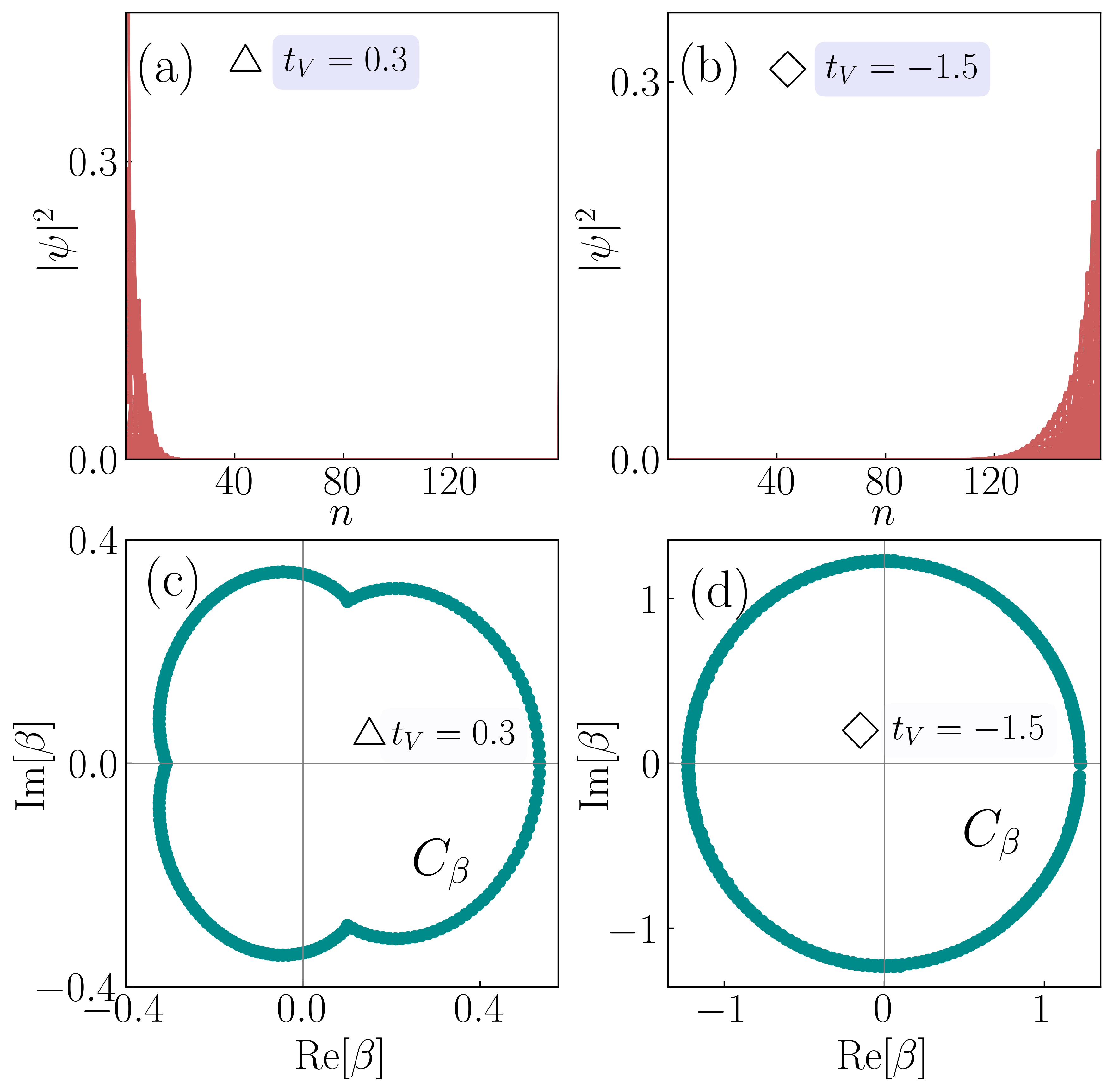}
\caption{Panels (a) and (b) display the probability distributions of the eigenstates for two specific values of \(t_V\), namely, \(t_V = 0.3\) and \(t_V = -1.5\), as indicated in Fig.~\ref{fig:2}(a) by the \(\triangle\) and \(\lozenge\)
symbols, respectively. The corresponding Floquet GBZs, denoted as \(C_\beta\), are shown in panels (c) and (d). These GBZ coordinates are then used to compute \(\nu_{0,\pi}\), presented in Fig.~\ref{fig:2}(d), which match accurately with the open boundary quasi-energy spectrum. The rest of the parameters are chosen as, $t_H=0.6$, $t_D=0.6$, $V_0=0.3$, $\gamma=0.4$, and $\omega=3$.}
\label{fig:3}
\end{figure}
\par Specifically, under a unitary (rotation) transformation $\mathcal{O}(t)$, the Floquet Hamiltonian is transformed as
\begin{equation}
H_{rot,k}(t) = \mathcal{O}(t)^{\dagger}(t)H_k(t)\mathcal{O}(t) - i\mathcal{O}^{\dagger}(t)\dot{\mathcal{O}}(t).
\end{equation}
For our analysis, we choose the rotating frame defined by
\begin{equation}
\mathcal{O}(t) = \exp\Big[-i (\hat n\!\cdot\!\boldsymbol\sigma) \frac{\omega t}{2}\Big],
\end{equation}
where $\hat{n} = (d_x/|d_k|,0,d_z/|d_k|)$. Additionally, the expressions for $d_x(k)$ and $d_z(k)$ are determined by isolating the coefficients of the $\sigma_x$ and $\sigma_z$ terms in Eq.~\eqref{eq:2}, evaluated at $\phi = \pi/2$. Working in this rotating frame simplifies the calculation by enabling a systematic expansion of the effective Hamiltonian in powers of the inverse driving frequency, an approach known as the Magnus expansion \cite{Benito2014,EckardtAnisimovas2015,Wang2020}. This yields the Floquet effective Hamiltonian as
\begin{equation}
H_{\text{eff}} = H_{rot,0}(k) + \sum_{m \neq 0} \frac{[H_{rot,0},H_m]}{m\omega} + \sum_{m>0} \frac{[H_m,H_{-m}]}{m\omega},
\label{magnus}
\end{equation}
where $H_{\pm m}$ denotes the Fourier components as defined earlier in Eq.~\eqref{eq:9}. 
Moreover, based on the preceding discussions, in the high-frequency regime, the expansion can be safely truncated at the first order, resulting in an effective Hamiltonian of the form
\begin{equation}
H_{\text{eff}} = d'_x(k) + d'_z(k),
\label{Floquet_Dirac}
\end{equation}
thus providing an approximate expression for the $d$-vectors required in constructing the Floquet GBZ, and are given as,
\begin{subequations}
    \begin{align}
        d_x^{\prime}(k) &= \left[ 1 - \frac{\omega}{2 |d_{k}|} \right] d_x(k) + \frac{V_0}{2|d_{k}|^2} d^2_z(k), \\
        d_z^{\prime}(k) &= \left[ 1 - \frac{\omega}{2 |d_{k}|} \right] d_z(k) - \frac{V_0}{2|d_{k}|^2} d_x (k) d_z (k).
    \end{align}
    \label{d_vectors_harmonic}
\end{subequations}
\begin{figure}[t]
         \includegraphics[width=\columnwidth]{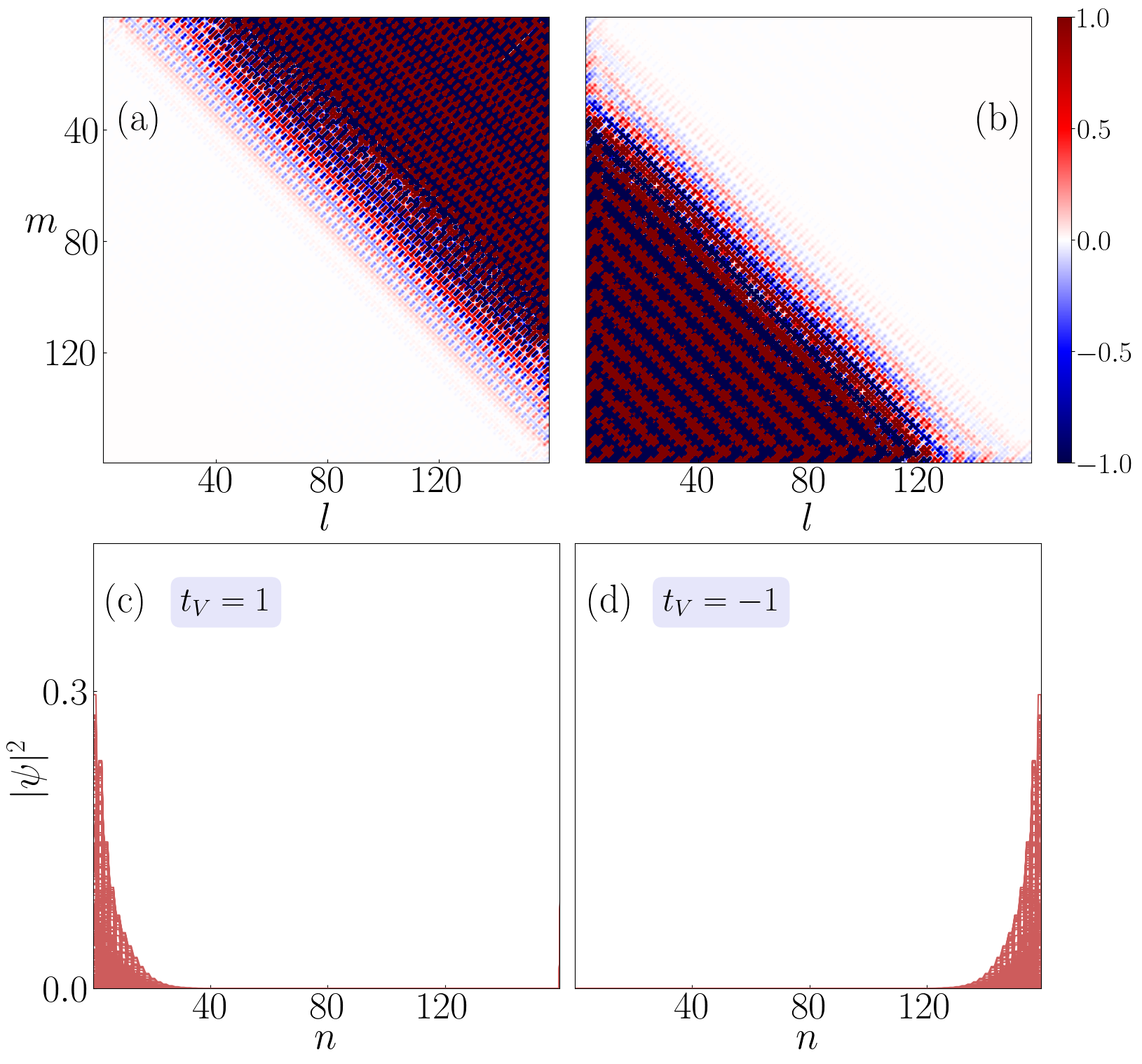}
\caption{The associated expansion coefficients corresponding to the Hamiltonian for $t_V=1$ and $t_V=-1$ in the coordinate basis have been shown in panels (a) and (b), respectively. Here, $l$ and $m$ denote the base indices, namely, the row and column indices of the Hamiltonian matrix in the coordinate basis. While panels (c) and (d) denote the distribution of the NH skin modes at the edges corresponding to positive ($t_V=1$) and negative ($t_V=-1$) values of $t_V$, respectively. The other parameters have been chosen as $t_H=t_D=0.6$, $V_0=0.3$, $\omega=1$, and $\gamma=0.4$.}
\label{fig:4}
\end{figure}
A detailed derivation of the expressions for these $d$-vectors is provided in the Supplemental Material (S1)~\cite{supp}. At this stage, it is also important to understand that before delving into the expansion, the unitary rotation was performed to transform the Hamiltonian into a rotating frame oscillating at $\omega/2$, effectively capturing single-photon absorption/emission processes. In this rotating frame, the rapidly oscillating terms are eliminated, and the remaining slow envelope dynamics are well approximated by a time-independent effective Hamiltonian. As a result, the phase boundaries obtained using these $d$-vectors can remain accurate not only in the strict high-frequency regime ($\omega \gg D$), but also provide good approximations in the intermediate regime ($\omega \sim D$). \par Next, we focus on the non-Bloch framework, which forms the basis for constructing the GBZ. In this approach, the Floquet Bloch Hamiltonian is reformulated in terms of the complex variable $\beta = e^{ik}$, where $k \in [-\pi, \pi]$, yielding the non-Bloch Floquet Hamiltonian $H(\beta)$. The allowed values of $\beta$ are obtained by solving the characteristic equation $\det[H(\beta) - E] = 0$, resulting in an algebraic equation of an even degree. While this method has been successfully applied to various static NH systems \cite{PhysRevLett.123.066404,YokomizoMurakami2021,Lin2023}, extending it to Floquet scenarios is technically challenging due to their inherent mathematical complexity. Unlike static systems, where the effective Hamiltonian typically contains only short-range couplings leading to a characteristic polynomial of a finite order in $\beta$, periodic driving induces effective long-range interactions. This results in a more intricate characteristic equation that complicates the direct construction of the GBZ. This is precisely why a high-frequency expansion was carried out earlier, as it yields the Hamiltonian in terms of approximate expressions for the $d$-vectors, making further analytical treatment feasible. We have provided a systematic and detailed derivation of the Floquet GBZ, obtained by analytically solving the characteristic polynomial formulated in terms of the $d$-vectors in the Supplemental Material (S1)~\cite{supp}.
\par Figs.~\ref{fig:3}(c) and \ref{fig:3}(d) illustrate the Floquet GBZ corresponding to the two representative points highlighted in Fig.~\ref{fig:2}(a), namely $t_V = 0.3$ and $t_V = -1.5$, respectively. It is clear from these plots that the GBZ consistently deviates from the conventional BZ, which underpins the emergence of NHSE. Notably, the radius of the GBZ encodes the localization direction of the states; that is, when all states are localized towards the left (right) edge of the chain, the radius of the GBZ becomes smaller (larger) than 1. Additionally, these GBZ profiles exhibit pronounced sharp features, commonly referred to as \textit{cusps}, appearing along their circumference. Such cusps typically arise when three out of the four solutions for $\beta$ attain the same absolute value. Interestingly, in the regime of weak driving strengths, the contributions from the second terms in Eqs.~\eqref{d_vectors_harmonic} become negligible. This simplification reduces the characteristic polynomial for $\beta$ to a quadratic form, implying that under weak driving conditions, the GBZ may approximate a circular trajectory with a constant radius, akin to the behavior observed in static systems.
\par Before delving into the characterization of distinct topological phases using the GBZ framework, it is important to recognize that the unique features exhibited by the GBZ stem from the pivotal role of periodic driving in effectively simulating long-range interactions within the system. This can be understood by observing that, depending on the sign of $t_V$, all the elements in the effective Hamiltonian ($H_{\text{eff}}$), corresponding to either the upper or lower diagonal sites, can become completely occupied, leaving the other set entirely empty. This behavior arises due to the interplay between periodic driving and non-Hermiticity, wherein the imaginary onsite potential introduces an effective non-reciprocity in the hopping amplitudes. Such a non-reciprocal effect is further amplified by the presence of the periodic drive, especially in the low-frequency ($T\gg1$) limit. As shown in Fig.~\ref{fig:4}, for positive values of $t_V$, the upper diagonal elements are fully occupied, leading to all the eigenstates being localized at the left edge of the system. Conversely, when $t_V$ is negative, the lower diagonal elements become completely occupied, resulting in all eigenstates being localized at the right edge.
\begin{figure}[t]
         \includegraphics[width=\columnwidth]{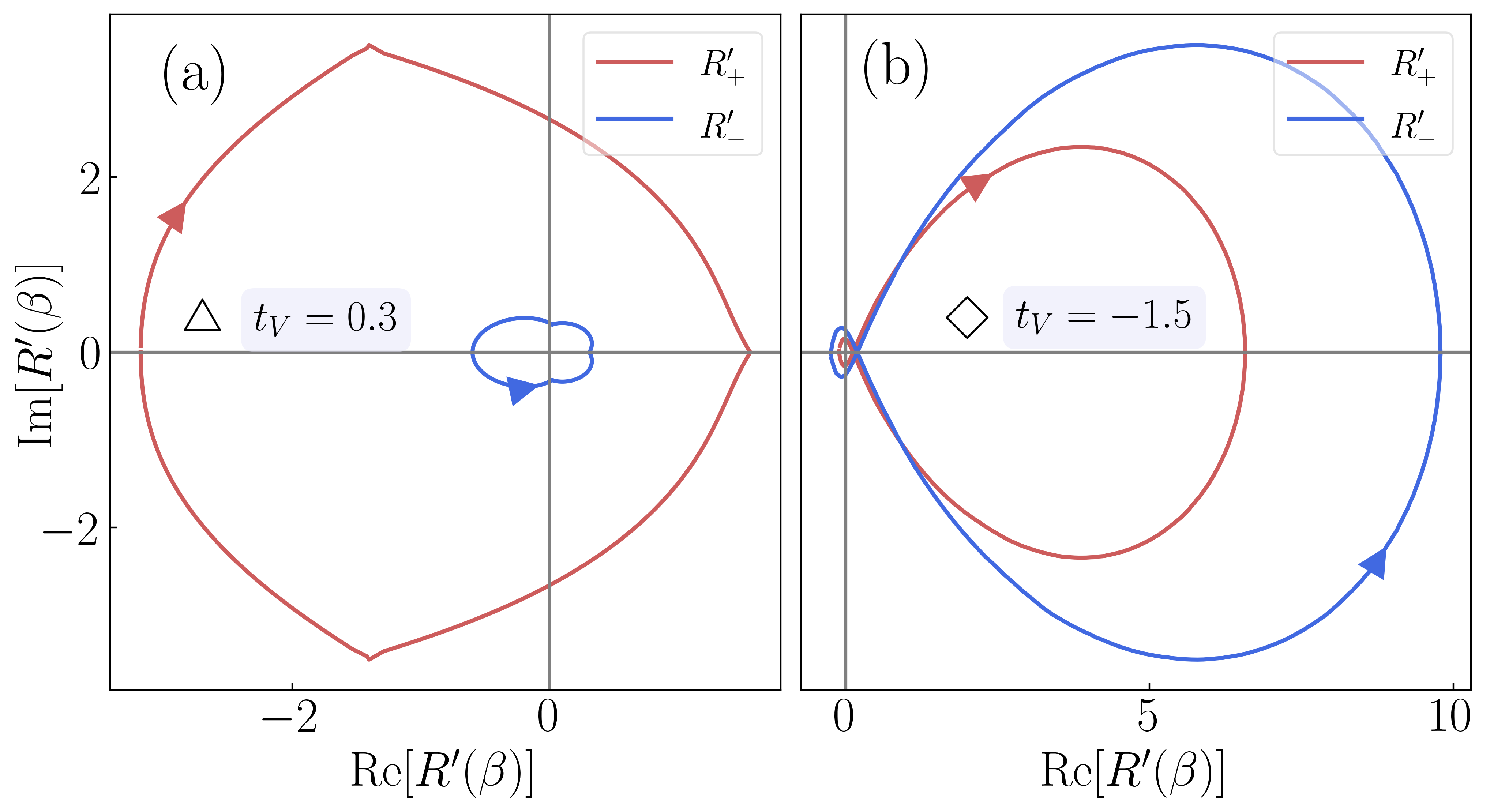}
\caption{Panels (a) and (b) display the trajectories of $R^{\prime}_{+}$ and $R^{\prime}_{-}$ on the complex plane along the GBZ for $t_V = 0.3$ (\(\triangle\)) and $t_V = -1.5$ (\(\lozenge\)), respectively. Notably, both the loops encircle the origin, indicating the presence of a topological phase. Furthermore, $R^{\prime}_{+}$ rotates in the clockwise direction, while $R^{\prime}_{-}$ rotates counterclockwise, resulting in a net finite winding of the $R$ vectors. All other parameters are the same as those used in Fig.~\ref{fig:2}.}
\label{fig:5}
\end{figure}
\par Finally, to conclude the discussion on non-Hermitian BBC, it is essential to accurately characterize the distinct topological phases. This can be effectively achieved by exploiting the chiral symmetry, which remains preserved even in the driven effective Hamiltonian., $H_{\text{eff}}$. Therefore, by substituting $e^{ik}$ with $\beta$, $H_{\text{eff}}$ can still be expressed in an off-diagonal matrix form after performing a suitable rotation of the spin basis $\sigma_y \rightarrow \sigma_z \rightarrow \sigma_y$. In this representation, the effective Hamiltonian takes the form,
\begin{equation}
H_{\text{eff}}(\beta) = R'_+(\beta) \sigma_{+} + R'_- (\beta)\sigma_{-},
\end{equation}
where
\begin{equation}
R'_{\pm}(\beta) = \left[ 1 - \frac{\omega}{2d_{\beta}} \mp \frac{V_0}{4 d^2_{\beta}} [R_+(\beta) - R_-(\beta)] \right] R_{\pm}(\beta),
\label{eq:20}
\end{equation}
and
\begin{equation}
R_{\pm} (\beta) = t_V \pm \gamma + t_D (\beta + \beta^{-1}) \mp t_H (\beta - \beta^{-1}).
\label{eq:21}
\end{equation}
Once these $R$-vectors are constructed, following the method outlined in Refs. \cite{PhysRevLett.123.066404, Xiao2020, PhysRevB.103.075126}, 
one can build the definition of non-Bloch invariant using the expression,
\begin{equation}
\label{Eq:23}
\mathcal{W}
=
\frac{i}{2\pi}
\oint_{C_{\beta}} dq\, q^{-1}(\beta)
=
-\frac{1}{4\pi}
\left[
\arg R'_+ (\beta)
-
\arg R'_- (\beta)
\right]_{C_{\beta}},
\end{equation}
where the integral is performed along the GBZ contour $C_{\beta}$.
The corresponding non-Bloch invariant is determined by tracking how the phases of $R^{\prime}_{\pm}$ evolve as $\beta$ traverses the GBZ ($C_{\beta}$) in the anti-clockwise direction. This behavior is illustrated in Fig.~\ref{fig:5}(a) and \ref{fig:5}(b), which depict the trajectories of $R^{\prime}_+$ and $R^{\prime}_-$ corresponding to the GBZs shown in Figs.~\ref{fig:3}(c) and \ref{fig:3}(d). Notably, in both cases, the loops encircle the origin, indicating the presence of a topological phase, consistent with the open boundary spectra displayed in Fig.~\ref{fig:2}(a). Further, in both the figures, \( R_+ \) rotates clockwise while \( R_- \) rotates counterclockwise, resulting in a finite value of \( \mathcal{W} \) due to the negative sign in Eq.~(\ref{Eq:23}). Notably, the loop traced by \( R_+ \) is larger than that of \( R_- \) in Fig.~\ref{fig:5}(a), whereas in Fig.~\ref{fig:5}(b), the situation is reversed, that is the \( R_- \) loop is larger than the \( R_+ \) loop.
\par While this approach confirms the existence of a topological phase, it does not distinguish whether the edge modes correspond to zero modes, $\pi$ modes, or a combination of both. Furthermore, simply counting the number of how many times these $R$-vectors wind around the origin does not yield the correct number of edge states, particularly when both zero and $\pi$ modes coexist. To address this limitation and accurately enumerate the distinct edge states, it becomes essential to employ a pair of invariants, which we derive using the concept of \textit{symmetric time frames} \cite{AsbothObuse2013,Asboth2014}. To begin with, it is useful to remember that Floquet eigenstates correspond to the eigenstates of the stroboscopic time evolution operator, $\hat{U}(T)$, which evolves the system over one complete period.
Interestingly, the symmetry properties of the effective Hamiltonian, $H_{\text{eff}}$ obtained from the stroboscopic evolution can depend on the chosen time frame within a period. Moreover, if we identify $t=T/2$ as a time reversal symmetric point then the entire cycle can be symmetrically divided into two parts. 
Let $F$ and $G$ denote the time evolution of the first and second part of the cycle respectively, that is,
\begin{equation}
    F = \mathcal{T} e^{-i \int_{0}^{t} H(t) dt} ~ ; ~ G = \mathcal{T} e^{-i \int_{t}^{T} H(t) dt}. 
\end{equation}
Consequently, these two segments are chiral symmetric partner to each other, and one can construct two symmetric time frame operators, namely,  $\hat{U}_1 = \hat{F} \hat{G}$ and $\hat{U}_2 = \hat{G} \hat{F}$, which not only reproduce the same quasi-energy spectrum as the original effective Hamiltonian, but also recover the symmetries associated with the static model that were otherwise lost in the driven scenario. 
Furthermore, depending on the periodic table of Floquet topological insulators \cite{RoyHarper2017}, each nontrivial phase in the system can be then characterized by a pair of winding numbers, $\nu_{0}$ and $\nu_{\pi}$ given as,
\begin{equation}
    \nu_0 = \frac{\nu_{1} + \nu_{2}}{2}, \quad \nu_{\pi} = \frac{\nu_{1} - \nu_{2}}{2},
\label{eq:winding_two_frames}
\end{equation}
where $\nu_{1}$ and $\nu_{2}$ denote the conventional definition of winding numbers \cite{Ryu2010,Schnyder2008} associated with the effective Hamiltonians in the two symmetric time frames, with the integration being carried out along GBZ contour $C_{\beta}$.
In Fig.~\ref{fig:2}(b), we present the variation of the two invariants, $\nu_0$ and $\nu_{\pi}$, as functions of $t_V$, which accurately captures the emergence of both zero and $\pi$ modes observed in Fig.~\ref{fig:2}(a). 
\begin{figure}[t]
         \includegraphics[width=0.9\columnwidth]{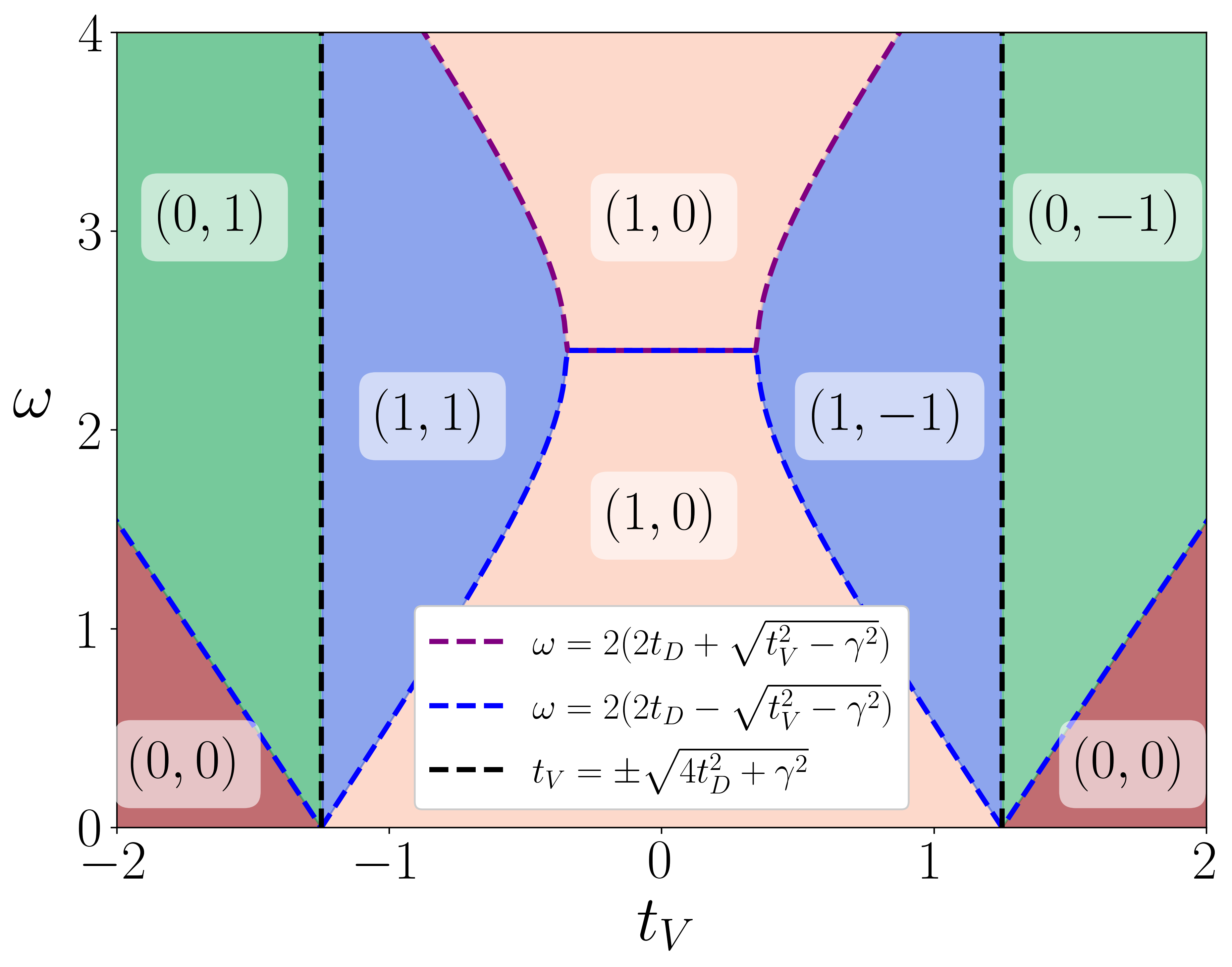}
\caption{Topological phase diagram in the $\omega$–$t_V$ plane characterized by the pair $(\nu_0, \nu_\pi)$. The phase boundaries are determined by solving the gap-closing conditions of the effective Hamiltonian $H_{\text{eff}}$ derived using high-frequency expansion. The other parameters are set as $t_H = t_D = 0.6$, $V = 0.3$, and $\gamma = 0.4$.
}
\label{fig:6}
\end{figure}
\par 
Notably, the computation of each invariant involves integrating over the Floquet GBZ, which we estimate through a high-frequency expansion of the Floquet Hamiltonian. Therefore, due to the technical complexities of these procedures, constructing a comprehensive topological phase diagram can turn out to be computationally cumbersome. Nevertheless, it is still possible to identify the distinct phase boundaries without explicitly considering both symmetric frames. Specifically, in the weak driving limit, the criterion for gap closing, that is, $R^{\prime}_+(\beta)R^{\prime}_-(\beta) = E^2(\beta) = 0$ can be obtained from the simultaneous conditions, $R_+(\beta) R_-(\beta) = 0$ and $\omega = 2 d_{\beta}$ (see Eq.~\eqref{eq:20} and Eq.~\eqref{eq:21}), leading to the phase boundary conditions given as
\begin{subequations}
\begin{equation}
    \omega = 2 | 2t_D \pm \sqrt{t_V^2 - \gamma^2}  |,
\end{equation}
\begin{equation}
    t_V = \pm \sqrt{4t_D^2 + \gamma^2},
\end{equation}
\end{subequations}
provided $t_D = t_H$ and the driving strength is sufficiently small compared to the bandwidth. Note that, a smaller driving strength is chosen to eliminate the quartic $\beta$-terms in Eq.~\eqref{eq:20}. Furthermore, the condition $t_D = t_H$ is imposed solely for analytical convenience. While one could in principle consider $t_H \neq t_D$, this does not introduce any new topological features apart from a possible shift in the phase boundaries. However, in that case, the conditions for gap closing, namely $R_+R_- = 0$ and $\omega = 2d_\beta$, would assume a much more complicated form. 
Fig.~\ref{fig:6} illustrates these phase boundaries in the $\omega$-$t_V$ plane with other parameters being fixed as in Fig.~\ref{fig:2}. Evidently, these three distinct boundaries divide the parameter space into eight sub-regions (as shown in Fig.~\ref{fig:6}). Furthermore, via analyzing the quasi-energy spectrum and evaluating the two invariants at representative points within these regions, one can classify each sub-region as a distinct non-trivial phase characterized by its corresponding pair of $\nu_0$ and $\nu_{\pi}$ values, as shown in Fig.~\ref{fig:6}. 
\begin{figure}[t]
         \includegraphics[width=\columnwidth]{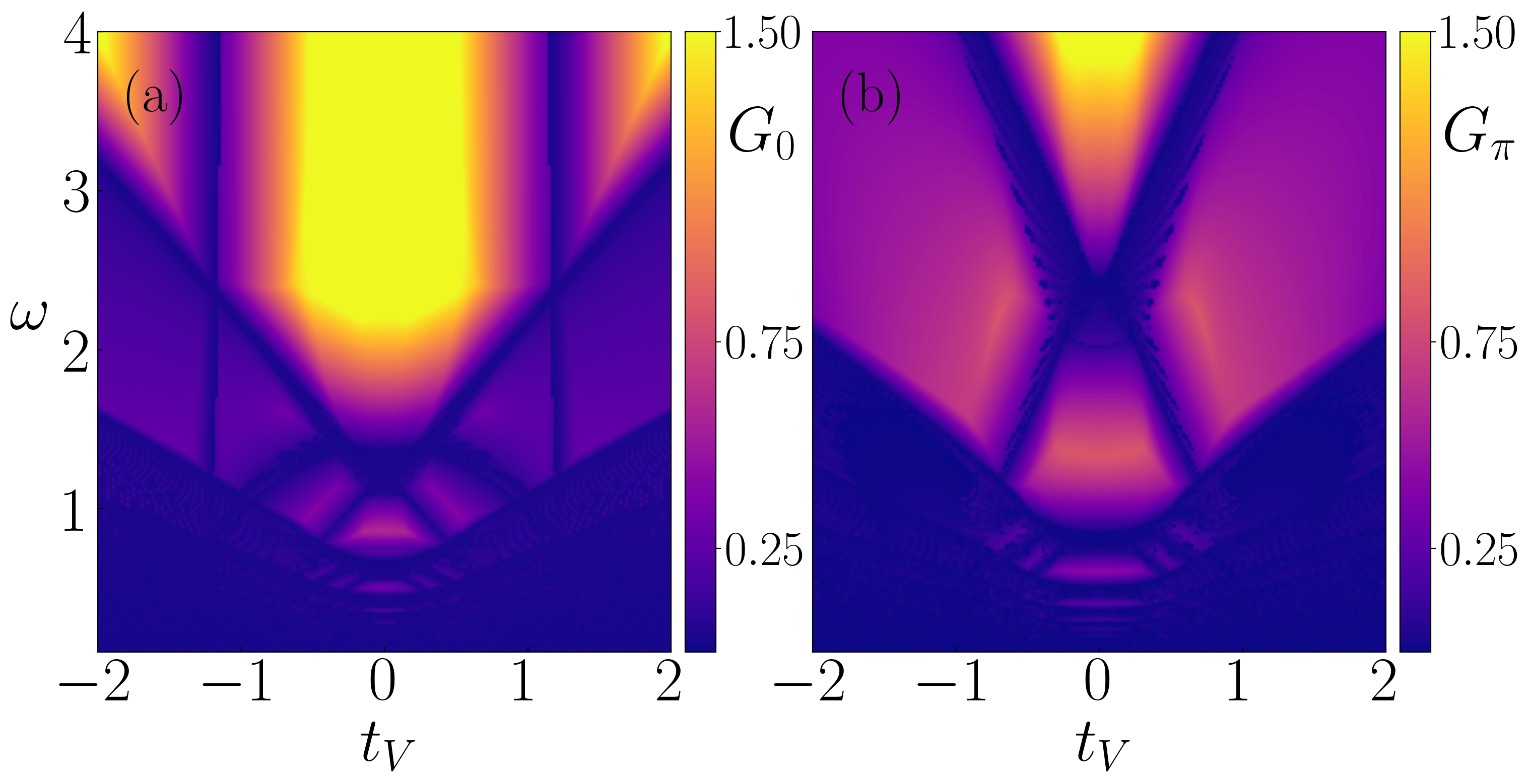}
\caption{The figure depicts the bulk gap invariant corresponding to both zero ($G_0$) and $\pi$ energy modes ($G_{\pi})$ in the $t_V$-$\omega$ plane. The $G_0$–$G_{\pi}$ diagram highlights regions in the low-frequency regime where the system effectively exhibits gapless states, indicating scenarios in which the topological invariants, as evaluated in Fig. \ref{fig:6}, may not provide a reliable characterization. The remaining parameters are chosen as $t_H=t_D=0.6$, $V_0 = 0.3$, and $\gamma=0.4$.}
\label{fig:7_new}
\end{figure}
\par At this stage, it is important to note that the phase diagram presented in the manuscript has been extended to the low-frequency regime, even down to values close to zero. While it is expected that the results obtained from the Magnus expansion may not remain quantitatively accurate in this regime, we have expressed the results corresponding to the complete frequency range to achieve a comprehensive understanding of the driven system. This is motivated by the fact that the crossover between the high- and low-frequency regimes is not universal but depends on the bandwidth, $D = 2(|t_V| + 2|t_D|)$, which varies with $t_V$. As a result, the boundary between different dynamical regimes shifts along the $t_V$ axis. Nevertheless, one should keep in mind that at very small $\omega$, the results still might fail. This can be understood physically by the fact that at very small $\omega$, the Floquet Brillouin zone gets compressed, and the system remains essentially gapless, leading to ill-defined topological invariants. To illustrate this, we have computed the bulk gaps corresponding to both zero and $\pi$ modes, as shown in Fig.~\ref{fig:7_new}. From this figure, one can clearly identify the frequency regimes where the system becomes gapless. Precisely these are the regions where the topological invariants inferred from Fig.~\ref{fig:6} may no longer be valid. Interestingly, the gapless boundaries associated with the zero and $\pi$ modes closely follow the topological phase boundaries in Fig.~\ref{fig:6}, with small deviations that stem from neglecting higher-order terms in Eq.~\eqref{eq:20}, in the weak driving assumption. Therefore, by comparing the bulk-gap diagram with the \textit{Magnus expansion} based phase diagram, one can systematically identify the range of frequencies where our analytical results remain quantitatively reliable. Thereby, building on this theoretical framework, in the next section, we explore an experimental realization using classical circuit platforms.
\begin{figure*}[t!]
    \centering
    \includegraphics[width=\linewidth, height=0.7\linewidth]{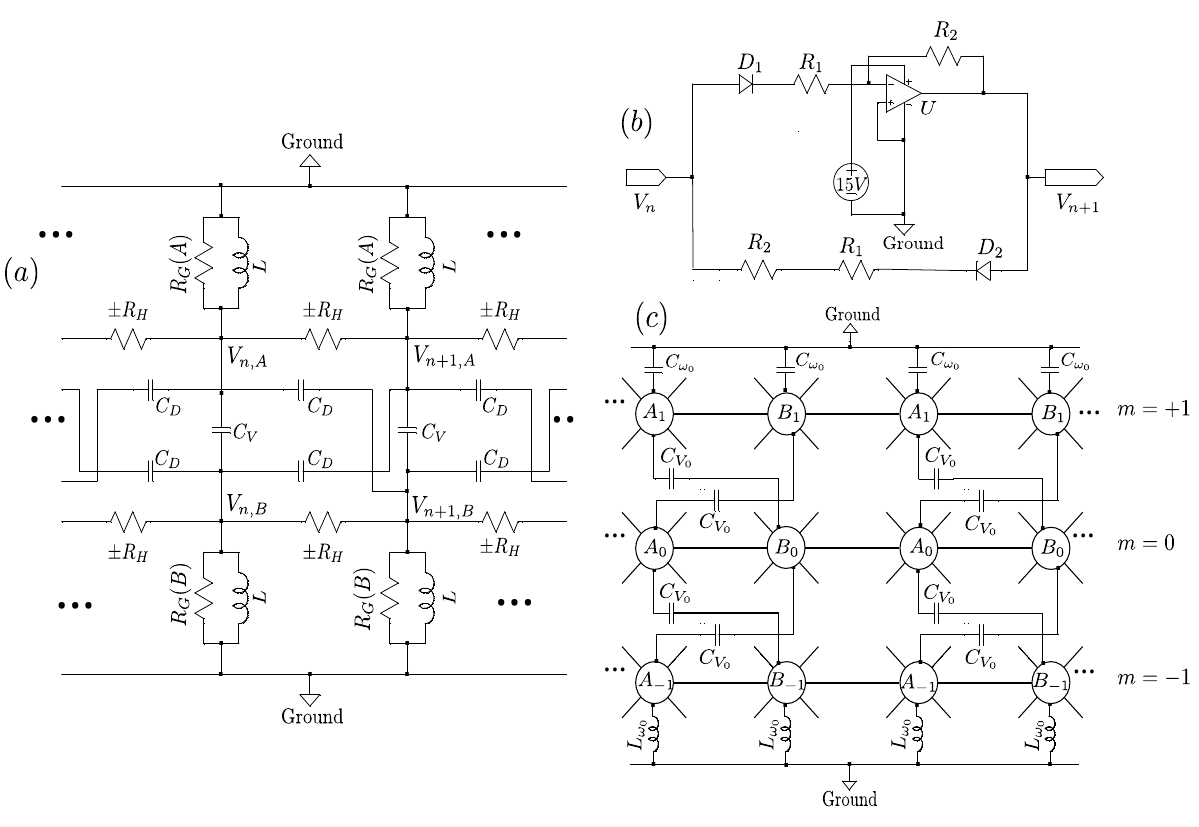}
    \caption{(a) The TEC diagram corresponding to the NH Creutz ladder. The output voltage at the subnode $A/B$ of the $n^{\text{th}}$ node is denoted by $V_{n, A/B}$. The circuit elements $\pm R_H$, $C_D$, and $C_V$ represent the hopping amplitudes $\pm i t_H$, $t_D$, and $t_V$, respectively. Further, the resistors $R_G(A/B)$ encode the NH onsite potentials $\pm i \gamma$. A grounded inductor $L$ is included at each subnode to tune the diagonal terms of the TEC Laplacian at the resonating frequency, $\omega_R$. (b) Circuit implementation of $\pm R_H$ between two adjacent nodes $n$ and $n+1$. A unidirectional current passing through diode $D_1$, from node $n$ to $n+1$, experiences an effective negative resistance of $-(R_1 + R_2)=-R_H$ due to the $\pi$ phase shift in the output voltage $V_{n+1}$, generated by the inverting amplifier configuration of the op-amp $U$. However, while flowing in the reverse direction, that is, from $n+1$ to $n$, the current experiences a positive resistance $(R_1 + R_2)=R_H$. (c) The TEC diagram corresponding to the NH Creutz ladder under a harmonic drive applied to the vertical hopping. The diagram illustrates three replicas of the original TEC, corresponding to Floquet indices $|m| = 0, 1$, interconnected by capacitors $C_{V_0}$. These capacitors simulate the Floquet drive by coupling adjacent Floquet layers.}
    \label{fig:7}
\end{figure*}

\section{\label{sec:level4}Topolectrical circuit of a Creutz ladder}

Thus far, we have theoretically demonstrated that the BBC can be restored in driven NH systems by characterizing each distinct phase through the non-Bloch invariants evaluated in two symmetric time frames.
Specifically, Figs.~\ref{fig:4}(c) and \ref{fig:4}(d) illustrate the appearance of NHSE arising from a staggered onsite imaginary potential, while Fig.~\ref{fig:6} presents the phase diagram indicating the presence or absence (including coexistence) of zero and $\pi$ modes in the driven TB model.
We now aim to design a TEC capable of faithfully mimicking the Floquet NH Creutz ladder.
The goal is not merely to simulate a quantum phenomenon in an electrical setting, but to directly visualize the localization behavior predicted by the TB model within the TEC framework.
The voltage and impedance profiles (IPs), being the key measurable quantities in such circuits, will serve as direct probes of the TB model's features.
Before proceeding, we briefly outline the formalism underlying the TEC implementation.

Similar to the Hamiltonian of a TB model, electrical circuit networks operate based on their Laplacians, which govern the network's response at each node \cite{F_Y_Wu_2004}.
For an electrical network with $N_0$ nodes, let $\mathcal{L}$ represent the Laplacian, and $V_k$ and $I_k$ denote the voltage and the total current through an external source at the $k^{\text{th}}$ node.
According to Kirchhoff's law, the following relation holds,
\begin{equation}
I_k=\sum_{p(k\ne p)}^{N_0}X_{kp}(V_k-V_p)+X_kV_k\quad\text{for}\quad k=1, 2, 3, \ldots, N_0,
\label{eq:current}
\end{equation}  
where $X_{kp}$ is the conductance between the $k^{\text{th}}$ node and the $p^{\text{th}}$ node.
Note that $X_{kk}$ has no physical meaning and is set to zero, while $X_k$ represents the resultant conductance between $k^{\text{th}}$ node and the ground.
With these definitions, Eq.~\eqref{eq:current} can be expressed as $I=\mathcal{L}V$, where $\mathcal{L}$ is the $N_0\times N_0$ Laplacian matrix with elements, $\mathcal{L}_{kp}=-X_{kp}+\delta_{kp}W_k,$ where $W_k=\sum_{p}X_{kp}+X_k.$
Now, the impedance between two nodes, namely $j$ and $k$, is given by,
\begin{equation}
	Z_{jk}=\sum_{q_n\ne0}\frac{|\Psi_{n,j}-\Psi_{n,k}|^2}{q_n},
	\label{eq:im}
\end{equation}
where $q_n$ is the $n^{\text{th}}$ eigenvalue of the Laplacian $\mathcal{L}$ and $\Psi_{n,j}$ is the $j^{\text{th}}$ element of the corresponding eigenmode.
This expression will help us obtain the IP of the TEC in subsequent sections.

\subsection{TEC construction}
To emulate the Creutz ladder, whose Hamiltonian is given by Eq.~\eqref{eq:2}, we construct an equivalent electrical circuit, shown in Fig.~\ref{fig:7}(a), whose Laplacian is denoted by $\mathcal{L}$.
Positive hopping amplitudes are realized using capacitors, contributing a positive imaginary admittance without dissipation under AC driving.
In particular, the diagonal (vertical) hopping amplitude $t_D$ $(t_V)$ in Fig.~\ref{fig:1} is implemented via a capacitor $C_D$ $(C_V)$ as shown in Fig.~\ref{fig:7}(a).
The admittance of $C_D$ ($C_V$) is given by $i\omega_R C_D$ $(i\omega_R C_V)$ and must satisfy the condition, 
\begin{equation}
|i\omega_R C_D| = t_D\quad\text{and}\quad|i\omega_R C_V| = t_V.
\label{eq:cdv}
\end{equation}
The value of the grounded inductor, $L$, is chosen to satisfy the following relation,
\begin{equation}
    i\omega_R (C_V+2C_D)+\frac{1}{i\omega_R L}=0,
\label{eq:L}
\end{equation}
at the resonating frequency, $\omega_R$.
The remaining terms in $H$ (see Eq.~\eqref{eq:2}) correspond to horizontal hopping modified by the Peierls phase $t_H e^{\pm i\phi}$, with $\phi = \pi/2$.
These terms are implemented using resistors, which introduce dissipative impedance into the circuit.
Note that the values must satisfy the condition
\begin{equation}
\left|\mp\frac{i}{R_H}\right| = |\pm i t_H|.
\label{eq:rh}
\end{equation}
This negative resistance, $-R_H$, is implemented using an inverting op-amp configuration, which functions as an amplifier with fixed negative gain, producing an output voltage of opposite polarity.
In Fig.~\ref{fig:7}(b), if $V_n$ ($V_{n+1}$) is taken as the input (output) voltage, assuming current flows only from $V_n$ to $V_{n+1}$, the effective impedance is given by $-(R_1 + R_2)$, where $R_1+R_2=R_H$.
This is equivalent to $-it_H$ in the TB model.
The configuration is implemented by placing a diode ($D_1$) at the inverting output of the op-amp ($U$), ensuring unidirectional current flow from the $n^{\text{th}}$ to the $(n+1)^{\text{th}}$ node, which effectively generates negative resistance for currents in the intended direction.
To realize the hopping term $+it_H$ (the Hermitian conjugate of $-it_H$), a parallel path is added between the $n^{\text{th}}$ and $(n+1)^{\text{th}}$ nodes, consisting of a series connection of $R_1$, $R_2$, and a diode ($D_2$) oriented opposite to $D_1$, as shown in Fig.~\ref{fig:7}(b).
To realize $\pm i\gamma$, given via Eq.~\eqref{eq:Ham4}, within the TEC we place resistors $R_G(A)$ and $R_G(B)$ at the subnodes $A_n$ and $B_n$, respectively, such that their magnitudes are equal but their signs are opposite, consistent with the relation $\gamma_A = -\gamma_B = \gamma$.
Thus, depending upon the sublattice index ($a_n$ or $b_n$) as illustrated in Fig.~\ref{fig:1}, the corresponding resistors must exhibit positive or negative resistance ($A_n$ or $B_n$), as shown in Fig.~\ref{fig:7}(a).
The values of $R_G(A)$ and $R_G(B)$ is given by by
\begin{equation}
R_G(A) = -R_G(B) = \frac{1}{|\gamma|}.
\label{eq:rg}
\end{equation}
The negative resistance, $-R_G(B)$, is implemented using the same configuration as $-R_H$.
The analytical details behind the formulation of the Laplacian and its equivalence to the Hamiltonian in Eq.~\eqref{eq:2} at the resonant angular frequency, $\omega_R$, are thoroughly discussed in the Supplemental Material (S3)~\cite{supp}.

We now turn our attention to the simulation of the Floquet Hamiltonian using TECs.
The design consists of multiple interconnected circuit layers, each representing a specific Fourier mode, as presented in Ref.~\cite{10.1063/5.0150118} for the case of a driven Hermitian SSH TEC.
Nevertheless, one can still opt for a real-time implementation with dynamically varying circuit parameters as suggested in Refs. \cite{PhysRevLett.119.093901,PhysRevApplied.18.054034}, which offer a direct way to probe Floquet physics.
In the context of the driven NH Creutz ladder, the mathematical form of the drive associated with the vertical hopping $t_V$ ensures that only the Fourier components with $m = 0$ and $|m| = 1$ contribute significantly to the Floquet expansion.
In Fig.~\ref{fig:7}(c), we schematically present the circuit diagram corresponding to the driven NH Creutz ladder.
The diagonal blocks $H \pm \omega$ in Eq.~\eqref{eq:9} are implemented using grounded inductors ($L_{\omega}$) and capacitors ($C_{\omega}$) connected to each subnode, depending on the sign of the shift, and the following relations must be satisfied,
\begin{equation}
|i\omega_R C_{\omega}| = \frac{1}{|i\omega_R L_{\omega}|} = \omega, \quad \text{and} \quad |i\omega_R C_{V_0}| = V_0.
\label{eq:c}
\end{equation}
The non-zero off-diagonal couplings between adjacent Floquet replicas ($|m|=0, 1$) are realized using capacitors of value $C_{V_0}$, corresponding to the driving amplitude $V_0$.
Thus, Fig.~\ref{fig:7}(c) provides the complete TEC-based realization of the driven NH Creutz ladder.
For convenience, we fix the following parameter values throughout the rest of the paper, $C_D=0.6\,\mu\text{F}\;(t_D=0.6),\;C_{\omega}=3\,\mu\text{F}\;(\omega=3),\;C_{V_0}=0.3\,\mu\text{F}\;(V_0=0.3),\;R_H=\frac{5}{3}\,\Omega\;(t_H=0.6)\;\text{and}\;R_G(A)=2.5\,\Omega\;(\gamma=0.4).$
Note that in TECs, imperfections inevitably arise due to the non-ideal characteristics of circuit components such as capacitors, inductors, resistors, and op-amps.
Examples include the parasitic resistance of capacitors and inductors, thermal noise in resistors, and the finite gain–bandwidth of op-amps \cite{Helbig2020}, all of which can cause deviations between experimental observations and theoretical predictions.

\subsection{Results and Analysis in TEC setting}

\begin{figure}[b]
    \centering
    \includegraphics[width=\linewidth]{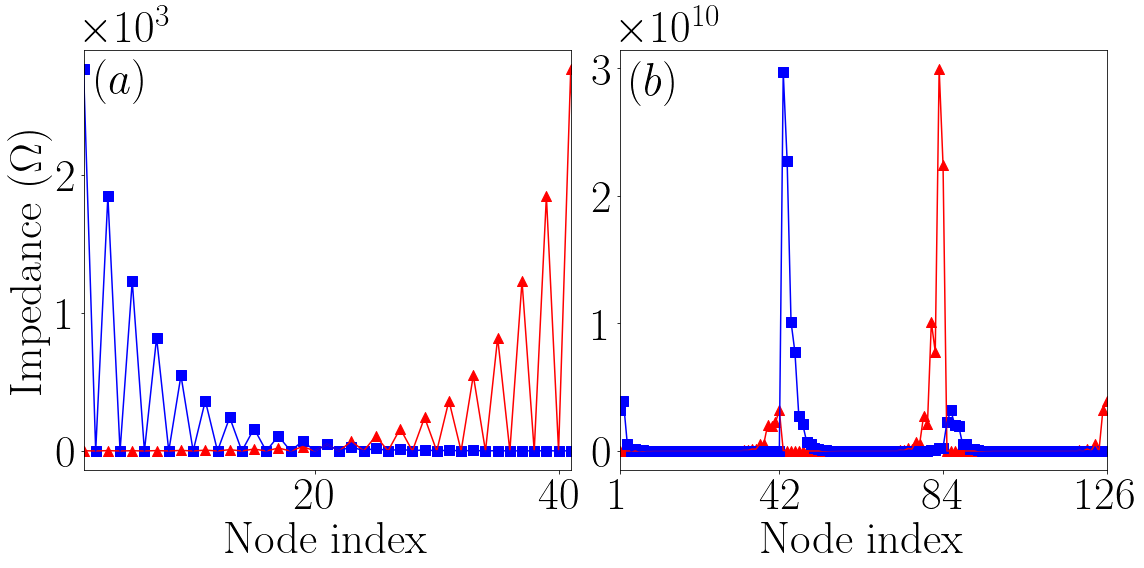}
    \caption{(a) The IP of the Hermitian Creutz ladder TEC, consisting of $21$ unit cells, is shown for both the trivial and topological regimes. The first and second edge IPs are represented by red triangles and blue squares, respectively. (b) The IP of the driven Creutz TEC, incorporating three Floquet replicas ($|m| = 0, 1$), is presented. In this case, the first edge IP is predominantly localized near the right edge of the $m = 0$ ladder ($84^{\text{th}}$ node), while the second edge IP localizes near its left edge ($43^{\text{rd}}$ node). Additionally, each edge IP exhibits minor localization near the adjacent edges to its primary localization node. The topological phase, which is evident from the presence of edge modes in the IP, is achieved by setting $C_V = 0.4\,\mu\text{F} < 2C_D$.}
    \label{fig:8}
\end{figure}

Let us first analyze the TEC corresponding to the static Hermitian Creutz ladder.
For the Hermitian model, the topological phase transition is governed by the vertical and diagonal hoppings in Eq.~\eqref{eq:2}, in particular by the ratio $\frac{t_V}{2t_D}$.
As shown in Fig.~\ref{fig:7}(a), setting $R_G(A)=R_G(B)\rightarrow\infty$ effectively removes the connection, yielding $\gamma=0$.
To get the zero-energy edge modes in our TEC setup, we calculate the IP of the Hermitian Creutz ladder, as shown in Fig.~\ref{fig:8}(a), using Eq.~\eqref{eq:im}.
In the topological regime, that is, for $C_V<2C_D$, two topological edge modes emerge, resulting in a substantial increase in the circuit impedance.
This manifests as two prominent edge IPs.
The first port is fixed at the first (last) node, while the second port is swept across all nodes to obtain the second (first) edge IP.
Each node is connected to two $C_D$ capacitors and one $C_V$ capacitor, and for a fixed amount of charge $Q$, this asymmetry leads to large potential differences.

Let $V_1$ and $V_2$ be the voltage differences across the vertical and diagonal capacitors, respectively, and satisfy the condition, $Q = C_V V_1 = 2C_D V_2$.
For $C_V < 2C_D$, this implies that the voltage drop across $C_V$ is greater than that across each $C_D$.
This further leads to $V_2/V_1 = C_V/(2C_D) < 1$ in the topological phase.
When the circuit is driven by an AC source, the potentials $V_1$ and $V_2$ oscillate out of phase, giving rise to a spatial voltage configuration of the form, $V(n) \propto (1, 0, -\frac{V_2}{V_1}, 0, (\frac{V_2}{V_1})^2, 0, -(\frac{V_2}{V_1})^3, 0, \hdots (-\frac{V_2}{V_1})^n, 0)$, where the index $n$ runs through the two-node unit cells \cite{Lee2018}.
This alternating pattern is reflected in the IP shown in Fig.~\ref{fig:8}(a).
For the driven case, the IP of the Hermitian Creutz TEC is constructed according to Fig.~\ref{fig:7}(c), with the values of the circuit elements kept unchanged.
As shown in Fig.~\ref{fig:8}(b), the IP exhibits localization at the edges of each Floquet replica ($|m| = 0, 1$).
Specifically, the first (second) edge IP is predominantly localized at the left (right) edge of the central Floquet replica ($m = 0$), consistent with the spatial probability distribution of the topological edge modes in the corresponding TB model.

\begin{figure}
    \centering
    \includegraphics[width=\linewidth]{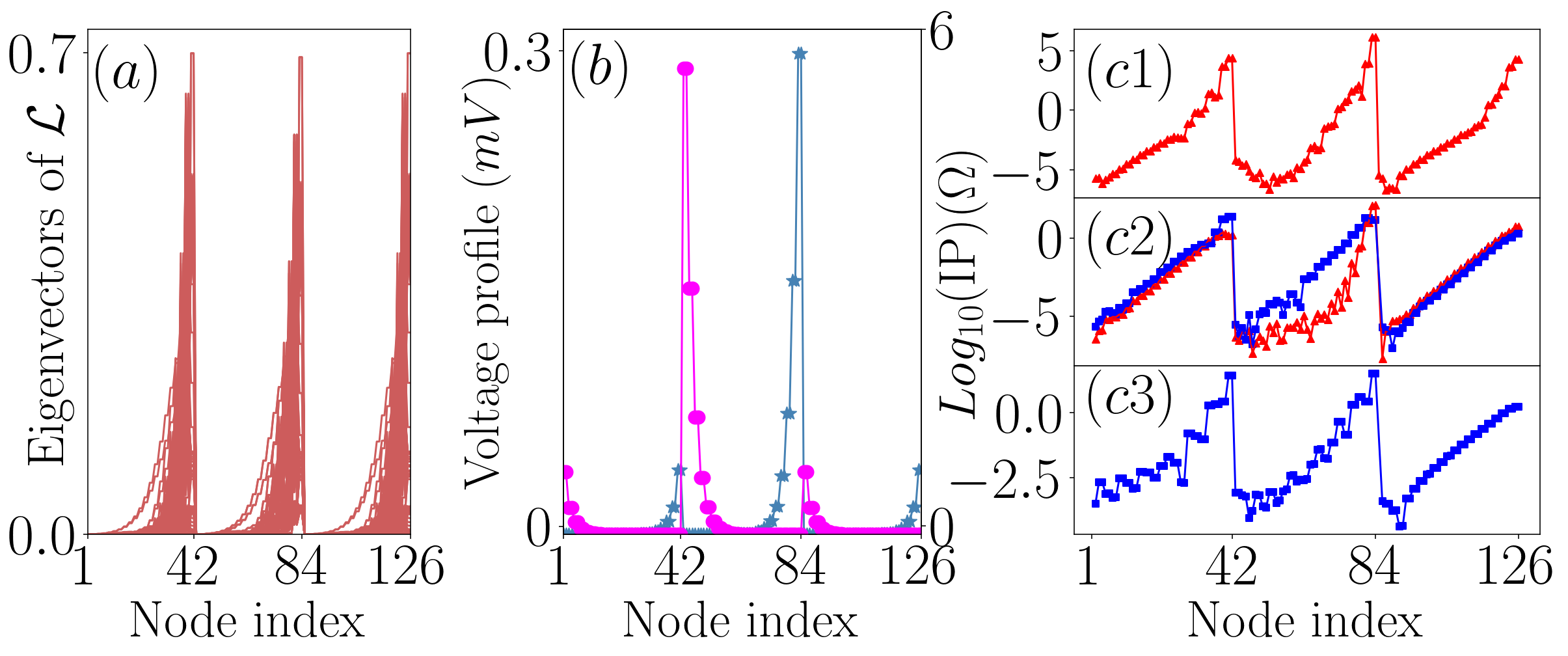}
    \caption{(a) The eigenmodes of the Laplacian for the driven NH Creutz ladder TEC with $R_G(A)=2.5\,\Omega$ exhibit localization near the edges of the Floquet replicas, signaling the emergence of NHSE in the TEC. The value of $C_V$ is chosen to be $0.3\,\mu\text{F}$. (b) A random node (node $66$ in this case) is excited using a current source of amplitude $1\,\mu A$, and the resulting voltage profile is plotted against the node index. The circles (peak at the $43^{\text{rd}}$ node) and stars (peak at the $84^{\text{th}}$ node) correspond to the cases with capacitors ($C_V = 1\,\mu\text{F}$) and inductors ($L_V = 1\,\mu\text{H}$), respectively. The IP in the $log_{10}$ scale are plotted for three different cases, (c1) $C_V=0.3\,\mu\text{F}$, (c2) $C_V=0.7\,\mu\text{F}$, and (c3) $C_V=1.7\,\mu\text{F}$. The red triangles (blue squares) correspond to the IP for the realization of the zero ($\pi$) energy modes via the TEC.}
    \label{fig:9}
\end{figure}

To construct the NH Floquet Creutz TEC, we include the resistors $R_G(A/B)$, with $R_G(A)=-R_G(B)=2.5\,\Omega$.
As demonstrated via Fig.~\ref{fig:4} in Sec.~\ref{sec:level3}~A, the staggered imaginary onsite potential $\pm i\gamma$ induces the NHSE in the driven NH Creutz ladder.
This phenomenon manifests in the TEC as shown in Fig.~\ref{fig:9}(a), where the absolute values of all the eigenmodes of the Laplacian are plotted against the node index.
The localization at the edges clearly indicates the presence of NHSE via theoretical simulations of the TEC setup.
However, in practical experiments, one cannot directly observe the eigenmodes of the Laplacian.
Instead, the measurable quantities are the impedance or the voltage at each node.
To experimentally probe NHSE in such a setup, one must excite the TEC at a randomly chosen node and measure the resulting voltage profile.
Fig.~\ref{fig:9}(b) illustrates this situation, where the $66^{\text{th}}$ node is excited using a current source, and the voltage at each node is recorded accordingly.
As demonstrated earlier, reversing the sign of $t_V$ alters the direction of the skin effect, resulting in the localization of bulk modes at the opposite edge of the system (see Figs.~\ref{fig:4}(c) and \ref{fig:4}(d)).
To implement this sign reversal of $t_V$ within the TEC, one must effectively reverse the sign of the vertical hopping, which requires replacing the capacitor $C_V$ with an inductor $L_V$ of the same magnitude.
The value of $L_V$ must satisfy the relation
\begin{equation*}
\frac{1}{|i\omega_RL_V|}=t_V,
\end{equation*}
and a corresponding adjustment must be made to the grounded inductor $L$.
It has been previously demonstrated that an onsite potential can induce a skin effect even in the absence of non-reciprocity in static cases \cite{PhysRevResearch.2.023265} via TECs.
This clearly demonstrates that one can control the direction and amplitude of voltage localization in the TEC by switching between vertical capacitors and inductors, an analogue of which has been elaborately discussed in the TB model.

We now explore the topological zero and $\pi$-modes, as illustrated in Fig.~\ref{fig:6} and discussed in Sec.~\ref{sec:level3}~B.
To demonstrate this, we consider three representative points from distinct regions of the phase diagram in Fig.~\ref{fig:6}, characterized by the following winding numbers, $1.\,\nu_0 = 1, \nu_\pi = 0,\;2.\,\nu_0 = 1, \nu_\pi = -1,\;3.\,\nu_0 = 0, \nu_\pi = -1.$
For the first case, corresponding to $\nu_0 = 1$, the logarithm of the IP is plotted in Fig.~\ref{fig:9}(c1). The impedance exhibits peaks at the right edges of each circuit row, indicating the presence of zero-energy modes.
In the second case ($\nu_0 = 1,; \nu_\pi = -1$), both zero and $\pi$ modes are predicted by the TB model.
However, since the $\pi$ modes have nonzero energies $\pm\frac{\omega}{2}$, they do not lead to impedance divergence in the conventional setup used to detect zero modes.
As a result, the $\pi$ modes are not directly visible in the standard IP measurement.
To detect the $\pi$-modes in the TEC, an effective chemical potential shift of $\mp\frac{\omega}{2}$ must be introduced.
This can be achieved by incorporating additional capacitors (to add $+\frac{\omega}{2}$) or inductors (to add $-\frac{\omega}{2}$)~\cite{10.1063/5.0150118}.
This adjustment ensures that the circuit impedance diverges at eigenenergies of the Laplacian corresponding to the $\pi$-mode energies of the TB model.
Accordingly, Fig.~\ref{fig:9}(c2) shows two configurations: one without added inductors (capturing the zero modes), and another with inductors that simulate the required chemical potential ($-\frac{\omega}{2}$), enabling the detection of the $\pi$-modes.
In both cases, the IP exhibits maxima at the right edges of the circuit rows, signaling the presence of both zero and $\pi$ modes in the TEC.
Finally, Fig.~\ref{fig:9}(c3) corresponds to the third region ($\nu_0 = 0,; \nu_\pi = -1$), where only $\pi$-modes are present, with no accompanying zero modes.
This is consistent with the phase diagram in Fig.~\ref{fig:6}.

\section{\label{sec:level6}conclusion}
To summarize, we have explored the intricate interplay between periodic driving and non-Hermitian effects in a Creutz ladder, thereby establishing a comprehensive theoretical framework based on the Floquet non-Bloch band theory.
By employing the Magnus expansion, we derive an approximate expression for the effective Floquet Hamiltonian, providing a formalism for evaluating the Floquet generalized Brillouin zone and analyzing the bulk-boundary correspondence for driven non-Hermitian systems.
We found that the conventional non-Bloch invariants derived from this effective Hamiltonian are insufficient to account for the full set of independent Floquet edge modes.
To resolve this, we introduced a symmetric frame approach that constructs two chiral-symmetric partner effective Hamiltonians.
The corresponding invariants associated with each partner, when combined appropriately, successfully identify and characterize the distinct edge modes, thereby restoring a modified bulk-boundary correspondence in the driven non-Hermitian model.

Recognizing the importance of experimental validation of our theoretical exploration, we have proposed a topolectrical circuit design that offers a practical and tunable platform for realizing both the skin modes and Floquet non-Hermitian topological phases.
Remarkably, the skin effect, realized via the voltage profile, persists even in the absence of non-reciprocity in the electrical setup, while the non-Hermitian Floquet topological phases manifest through the circuit's impedance profile.
Collectively, our work bridges two rapidly advancing fields, that is, Floquet engineering and non-Hermitian topology, and lays the groundwork for their unification within experimentally accessible platforms, thereby opening new avenues in the exploration of driven non-Hermitian systems.

\section{Data Availability}
The data that supports the findings of this article cannot be made publicly available.
The data are available upon reasonable request from the authors.\\

\section{Acknowledgments}
KR and DH acknowledge the research fellowship from the MoE, Government of India. Also, BT is supported in part by the Turkish Academy of Sciences (T\"{U}BA).\\

\bibliography{ref6}

@article{PhysRevLett.116.133903,
  title = {Anomalous {Edge} {State} in a {Non-Hermitian} {Lattice}},
  author = {Lee, Tony E.},
  journal = {Phys. Rev. Lett.},
  volume = {116},
  issue = {13},
  pages = {133903},
  numpages = {5},
  year = {2016},
  month = {Apr},
  publisher = {American Physical Society},
  doi = {10.1103/PhysRevLett.116.133903},
  url = {https://link.aps.org/doi/10.1103/PhysRevLett.116.133903}
}

@article{PhysRevLett.120.146402,
  title = {Topological {Band} {Theory} for {Non-Hermitian} {Hamiltonians}},
  author = {Shen, Huitao and Zhen, Bo and Fu, Liang},
  journal = {Phys. Rev. Lett.},
  volume = {120},
  issue = {14},
  pages = {146402},
  numpages = {6},
  year = {2018},
  month = {Apr},
  publisher = {American Physical Society},
  doi = {10.1103/PhysRevLett.120.146402},
  url = {https://link.aps.org/doi/10.1103/PhysRevLett.120.146402}
}

@article{Ghatak_2019,
title = {New topological invariants in non-{Hermitian} systems},
author = {Ananya Ghatak and Tanmoy Das},
journal = {Journal of Physics: Condensed Matter},
volume = {31},
issue = {26},
pages = {263001},
year = {2019},
month = {Apr},
publisher = {IOP Publishing},
doi = {10.1088/1361-648X/ab11b3},
url = {https://dx.doi.org/10.1088/1361-648X/ab11b3},
}

@article{PhysRevX.8.031079,
  title = {Topological {Phases} of {Non-Hermitian Systems}},
  author = {Gong, Zongping and Ashida, Yuto and Kawabata, Kohei and Takasan, Kazuaki and Higashikawa, Sho and Ueda, Masahito},
  journal = {Phys. Rev. X},
  volume = {8},
  issue = {3},
  pages = {031079},
  numpages = {33},
  year = {2018},
  month = {Sep},
  publisher = {American Physical Society},
  doi = {10.1103/PhysRevX.8.031079},
  url = {https://link.aps.org/doi/10.1103/PhysRevX.8.031079}
}

@article{PhysRevX.9.041015,
  title = {Symmetry and {Topology} in {Non-Hermitian Physics}},
  author = {Kawabata, Kohei and Shiozaki, Ken and Ueda, Masahito and Sato, Masatoshi},
  journal = {Phys. Rev. X},
  volume = {9},
  issue = {4},
  pages = {041015},
  numpages = {52},
  year = {2019},
  month = {Oct},
  publisher = {American Physical Society},
  doi = {10.1103/PhysRevX.9.041015},
  url = {https://link.aps.org/doi/10.1103/PhysRevX.9.041015}
}

@article{Ashida2020,
title = {{Non-Hermitian} {physics}},
author = {Yuto Ashida and Zongping Gong and Masahito Ueda},
journal = {Advances in Physics},
volume = {69},
issue = {3},
pages = {249-435},
year  = {2020},
month = {Apr},
publisher = {Taylor & Francis},
doi = {10.1080/00018732.2021.1876991},
url = {https://doi.org/10.1080/00018732.2021.1876991}
}

@article{RevModPhys.93.015005,
  title = {Exceptional topology of non-{Hermitian} systems},
  author = {Bergholtz, Emil J. and Budich, Jan Carl and Kunst, Flore K.},
  journal = {Rev. Mod. Phys.},
  volume = {93},
  issue = {1},
  pages = {015005},
  numpages = {31},
  year = {2021},
  month = {Feb},
  publisher = {American Physical Society},
  doi = {10.1103/RevModPhys.93.015005},
  url = {https://link.aps.org/doi/10.1103/RevModPhys.93.015005}
}

@article{PhysRevLett.121.086803,
  title = {Edge States and {Topological Invariants} of {Non-Hermitian Systems}},
  author = {Yao, Shunyu and Wang, Zhong},
  journal = {Phys. Rev. Lett.},
  volume = {121},
  issue = {8},
  pages = {086803},
  numpages = {8},
  year = {2018},
  month = {Aug},
  publisher = {American Physical Society},
  doi = {10.1103/PhysRevLett.121.086803},
  url = {https://link.aps.org/doi/10.1103/PhysRevLett.121.086803}
}

@article{PhysRevLett.121.026808,
  title = {Biorthogonal {Bulk-Boundary Correspondence} in {Non-Hermitian Systems}},
  author = {Kunst, Flore K. and Edvardsson, Elisabet and Budich, Jan Carl and Bergholtz, Emil J.},
  journal = {Phys. Rev. Lett.},
  volume = {121},
  issue = {2},
  pages = {026808},
  numpages = {6},
  year = {2018},
  month = {Jul},
  publisher = {American Physical Society},
  doi = {10.1103/PhysRevLett.121.026808},
  url = {https://link.aps.org/doi/10.1103/PhysRevLett.121.026808}
}

@article{PhysRevB.99.201103,
  title = {Anatomy of skin modes and topology in non-{Hermitian} systems},
  author = {Lee, Ching Hua and Thomale, Ronny},
  journal = {Phys. Rev. B},
  volume = {99},
  issue = {20},
  pages = {201103},
  numpages = {5},
  year = {2019},
  month = {May},
  publisher = {American Physical Society},
  doi = {10.1103/PhysRevB.99.201103},
  url = {https://link.aps.org/doi/10.1103/PhysRevB.99.201103}
}

@article{PhysRevLett.124.056802,
  title = {Non-{Hermitian Boundary Modes and Topology}},
  author = {Borgnia, Dan S. and Kruchkov, Alex Jura and Slager, Robert-Jan},
  journal = {Phys. Rev. Lett.},
  volume = {124},
  issue = {5},
  pages = {056802},
  numpages = {6},
  year = {2020},
  month = {Feb},
  publisher = {American Physical Society},
  doi = {10.1103/PhysRevLett.124.056802},
  url = {https://link.aps.org/doi/10.1103/PhysRevLett.124.056802}
}

@article{PhysRevLett.124.086801,
  title = {Topological {Origin} of {Non-Hermitian Skin Effects}},
  author = {Okuma, Nobuyuki and Kawabata, Kohei and Shiozaki, Ken and Sato, Masatoshi},
  journal = {Phys. Rev. Lett.},
  volume = {124},
  issue = {8},
  pages = {086801},
  numpages = {7},
  year = {2020},
  month = {Feb},
  publisher = {American Physical Society},
  doi = {10.1103/PhysRevLett.124.086801},
  url = {https://link.aps.org/doi/10.1103/PhysRevLett.124.086801}
}

@article{PhysRevB.97.121401,
  title = {Non-{Hermitian} robust edge states in one dimension: Anomalous localization and eigenspace condensation at exceptional points},
  author = {Martinez Alvarez, V. M. and Barrios Vargas, J. E. and Foa Torres, L. E. F.},
  journal = {Phys. Rev. B},
  volume = {97},
  issue = {12},
  pages = {121401},
  numpages = {6},
  year = {2018},
  month = {Mar},
  publisher = {American Physical Society},
  doi = {10.1103/PhysRevB.97.121401},
  url = {https://link.aps.org/doi/10.1103/PhysRevB.97.121401}
}

@article{Heiss_2012,
title = {The physics of exceptional points},
author = {W D Heiss},
journal = {Journal of Physics A: Mathematical and Theoretical},
volume = {45},
issue = {44},
pages = {444016},
year = {2012},
month = {oct},
publisher = {IOP Publishing},
doi = {10.1088/1751-8113/45/44/444016},
url = {https://dx.doi.org/10.1088/1751-8113/45/44/444016},
}

@article{PhysRevLett.118.040401,
  title = {Edge Modes, {Degeneracies}, and {Topological Numbers} in {Non-Hermitian Systems}},
  author = {Leykam, Daniel and Bliokh, Konstantin Y. and Huang, Chunli and Chong, Y. D. and Nori, Franco},
  journal = {Phys. Rev. Lett.},
  volume = {118},
  issue = {4},
  pages = {040401},
  numpages = {6},
  year = {2017},
  month = {Jan},
  publisher = {American Physical Society},
  doi = {10.1103/PhysRevLett.118.040401},
  url = {https://link.aps.org/doi/10.1103/PhysRevLett.118.040401}
}

@article{PhysRevLett.123.066404,
  title = {Non-Bloch Band Theory of Non-Hermitian Systems},
  author = {Yokomizo, Kazuki and Murakami, Shuichi},
  journal = {Phys. Rev. Lett.},
  volume = {123},
  issue = {6},
  pages = {066404},
  numpages = {6},
  year = {2019},
  month = {Aug},
  publisher = {American Physical Society},
  doi = {10.1103/PhysRevLett.123.066404},
  url = {https://link.aps.org/doi/10.1103/PhysRevLett.123.066404}
}

@article{El-Ganainy2018,
author={El-Ganainy, Ramy and Makris, Konstantinos G. and Khajavikhan, Mercedeh and Musslimani, Ziad H. and Rotter, Stefan and Christodoulides, Demetrios N.},
title={Non-{Hermitian} physics and {PT} symmetry},
journal={Nature Physics},
year={2018},
month={Jan},
day={01},
volume={14},
number={1},
pages={11-19},
issn={1745-2481},
doi={10.1038/nphys4323},
url={https://doi.org/10.1038/nphys4323}
}

@article{Eichelkraut2013,
author={Eichelkraut, T. and Heilmann, R. and Weimann, S. and St{\"u}tzer, S. and Dreisow, F. and Christodoulides, D. N. and Nolte, S. and Szameit, A.},
title={Mobility transition from ballistic to diffusive transport in non-{Hermitian} lattices},
journal={Nature Communications},
year={2013},
month={Sep},
day={26},
volume={4},
number={1},
pages={2533},
issn={2041-1723},
doi={10.1038/ncomms3533},
url={https://doi.org/10.1038/ncomms3533}
}

@article{PhysRevLett.114.173902,
  title = {Topological {Properties} of {Linear Circuit Lattices}},
  author = {Albert, Victor V. and Glazman, Leonid I. and Jiang, Liang},
  journal = {Phys. Rev. Lett.},
  volume = {114},
  issue = {17},
  pages = {173902},
  numpages = {6},
  year = {2015},
  month = {Apr},
  publisher = {American Physical Society},
  doi = {10.1103/PhysRevLett.114.173902},
  url = {https://link.aps.org/doi/10.1103/PhysRevLett.114.173902}
}

@article{Helbig2020,
author={Helbig, T. and Hofmann, T. and Imhof, S. and Abdelghany, M. and Kiessling, T. and Molenkamp, L. W. and Lee, C. H. and Szameit, A. and Greiter, M. and Thomale, R.},
title={Generalized bulk--boundary correspondence in non-{Hermitian} topolectrical circuits},
journal={Nature Physics},
year={2020},
month={Jul},
day={01},
volume={16},
number={7},
pages={747-750},
issn={1745-2481},
doi={10.1038/s41567-020-0922-9},
url={https://doi.org/10.1038/s41567-020-0922-9}
}

@article{Lee2018,
author={Lee, Ching Hua and Imhof, Stefan and Berger, Christian and Bayer, Florian and Brehm, Johannes and Molenkamp, Laurens W. and Kiessling, Tobias and Thomale, Ronny},
title={Topolectrical {Circuits}},
journal={Communications Physics},
year={2018},
month={Jul},
day={23},
volume={1},
number={1},
pages={39},
issn={2399-3650},
doi={10.1038/s42005-018-0035-2},
url={https://doi.org/10.1038/s42005-018-0035-2}
}

@article{Halder_2023,
doi = {10.1088/1361-648X/acadc5},
url = {https://dx.doi.org/10.1088/1361-648X/acadc5},
year = {2022},
month = {dec},
publisher = {IOP Publishing},
volume = {35},
number = {10},
pages = {105901},
author = {Halder, Dipendu and Ganguly, Sudin and Basu, Saurabh},
title = {Properties of the non-{Hermitian SSH} model: role of {PT} symmetry},
journal = {Journal of Physics: Condensed Matter}
}

@article{Halder_2024,
doi = {10.1088/1361-648X/ad4940},
url = {https://dx.doi.org/10.1088/1361-648X/ad4940},
year = {2024},
month = {may},
publisher = {IOP Publishing},
volume = {36},
number = {33},
pages = {335301},
author = {Halder, Dipendu and Basu, Saurabh},
title = {Parsing skin effect in a non-{Hermitian} spinless {BHZ}-like model},
journal = {Journal of Physics: Condensed Matter}
}

@article{PhysRevB.109.115407,
  title = {Circuit realization of a two-orbital non-{Hermitian} tight-binding chain},
  author = {Halder, Dipendu and Thomale, Ronny and Basu, Saurabh},
  journal = {Phys. Rev. B},
  volume = {109},
  issue = {11},
  pages = {115407},
  numpages = {10},
  year = {2024},
  month = {Mar},
  publisher = {American Physical Society},
  doi = {10.1103/PhysRevB.109.115407},
  url = {https://link.aps.org/doi/10.1103/PhysRevB.109.115407}
}

@article{Wang,
author = {Aoxi Wang  and Zhiqiang Meng  and Chang Qing Chen },
title = {Non-{Hermitian} topology in static mechanical metamaterials},
journal = {Science Advances},
volume = {9},
number = {27},
pages = {},
year = {2023},
doi = {10.1126/sciadv.adf7299},
URL = {https://www.science.org/doi/abs/10.1126/sciadv.adf7299}
}

@article{Fleury2015,
  title = {An invisible acoustic sensor based on parity-time symmetry},
  author = {Fleury, Romain and Sounas, Dimitrios and Alù, Andrea},
  journal = {Nat. Commun.},
  volume = {6},
  number = {1},
  year = {2015},
  doi = {10.1038/ncomms6905},
  url = {https://doi.org/10.1038/ncomms6905}
}

@article{PhysRevApplied.16.057001,
  title = {Controlling {Sound} in {Non-Hermitian Acoustic Systems}},
  author = {Gu, Zhongming and Gao, He and Cao, Pei-Chao and Liu, Tuo and Zhu, Xue-Feng and Zhu, Jie},
  journal = {Phys. Rev. Appl.},
  volume = {16},
  issue = {5},
  pages = {057001},
  numpages = {19},
  year = {2021},
  month = {Nov},
  publisher = {American Physical Society},
  doi = {10.1103/PhysRevApplied.16.057001},
  url = {https://link.aps.org/doi/10.1103/PhysRevApplied.16.057001},
}

@article{PhysRevB.99.161114,
  title = {Band structure engineering and reconstruction in electric circuit networks},
  author = {Helbig, Tobias and Hofmann, Tobias and Lee, Ching Hua and Thomale, Ronny and Imhof, Stefan and Molenkamp, Laurens W. and Kiessling, Tobias},
  journal = {Phys. Rev. B},
  volume = {99},
  issue = {16},
  pages = {161114},
  numpages = {5},
  year = {2019},
  month = {Apr},
  publisher = {American Physical Society},
  doi = {10.1103/PhysRevB.99.161114},
  url = {https://link.aps.org/doi/10.1103/PhysRevB.99.161114}
}

@article{PhysRevResearch.3.023056,
  title = {{Topolectric circuits: Theory and construction}},
  author = {Dong, Junkai and Juri\ifmmode \check{c}\else \v{c}\fi{}i\ifmmode \acute{c}\else \'{c}\fi{}, Vladimir and Roy, Bitan},
  journal = {Phys. Rev. Res.},
  volume = {3},
  issue = {2},
  pages = {023056},
  numpages = {22},
  year = {2021},
  month = {Apr},
  publisher = {American Physical Society},
  doi = {10.1103/PhysRevResearch.3.023056},
  url = {https://link.aps.org/doi/10.1103/PhysRevResearch.3.023056}
}

@article{PhysRevResearch.2.023265,
  title = {Reciprocal skin effect and its realization in a topolectrical circuit},
  author = {Hofmann, Tobias and Helbig, Tobias and Schindler, Frank and Salgo, Nora and Brzezi\ifmmode \acute{n}\else \'{n}\fi{}ska, Marta and Greiter, Martin and Kiessling, Tobias and Wolf, David and Vollhardt, Achim and Kaba\ifmmode \check{s}\else \v{s}\fi{}i, Anton and Lee, Ching Hua and Bilu\ifmmode \check{s}\else \v{s}\fi{}i\ifmmode \acute{c}\else \'{c}\fi{}, Ante and Thomale, Ronny and Neupert, Titus},
  journal = {Phys. Rev. Res.},
  volume = {2},
  issue = {2},
  pages = {023265},
  numpages = {11},
  year = {2020},
  month = {Jun},
  publisher = {American Physical Society},
  doi = {10.1103/PhysRevResearch.2.023265},
  url = {https://link.aps.org/doi/10.1103/PhysRevResearch.2.023265}
}

@article{7t7k-qg49,
  title = {Controlled probing of localization effects in the non-{Hermitian} {Aubry-Andr\'e} model via topolectrical circuits},
  author = {Halder, Dipendu and Basu, Saurabh},
  journal = {Phys. Rev. B},
  volume = {111},
  issue = {23},
  pages = {235447},
  numpages = {12},
  year = {2025},
  month = {Jun},
  publisher = {American Physical Society},
  doi = {10.1103/7t7k-qg49},
  url = {https://link.aps.org/doi/10.1103/7t7k-qg49}
}

@article{YANG20241,
title = {Circuit realization of topological physics},
journal = {Physics Reports},
volume = {1093},
pages = {1-54},
year = {2024},
note = {Circuit realization of topological physics},
issn = {0370-1573},
doi = {https://doi.org/10.1016/j.physrep.2024.09.007},
url = {https://www.sciencedirect.com/science/article/pii/S0370157324003302},
author = {Huanhuan Yang and Lingling Song and Yunshan Cao and Peng Yan}
}

@article{10.1063/5.0265293,
    author = {Sahin, Haydar and Jalil, Mansoor B. A. and Lee, Ching Hua},
    title = {Topolectrical circuits—{Recent} experimental advances and developments},
    journal = {APL Electronic Devices},
    volume = {1},
    number = {2},
    pages = {021503},
    year = {2025},
    month = {04},
    issn = {2995-8423},
    doi = {10.1063/5.0265293},
    url = {https://doi.org/10.1063/5.0265293}
}

@article{F_Y_Wu_2004,
doi = {10.1088/0305-4470/37/26/004},
url = {https://dx.doi.org/10.1088/0305-4470/37/26/004},
year = {2004},
month = {jun},
publisher = {},
volume = {37},
number = {26},
pages = {6653},
author = {F Y Wu},
title = {Theory of resistor networks: the two-point resistance},
journal = {Journal of Physics A: Mathematical and General}
}

@article{PhysRevB.92.085118,
  title = {Characterization of topological phase transitions via topological properties of transition points},
  author = {Li, Linhu and Chen, Shu},
  journal = {Phys. Rev. B},
  volume = {92},
  issue = {8},
  pages = {085118},
  numpages = {7},
  year = {2015},
  month = {Aug},
  publisher = {American Physical Society},
  doi = {10.1103/PhysRevB.92.085118},
  url = {https://link.aps.org/doi/10.1103/PhysRevB.92.085118}
}

@article{PhysRevB.83.245132,
  title = {{Inversion-symmetric topological insulators}},
  author = {Hughes, Taylor L. and Prodan, Emil and Bernevig, B. Andrei},
  journal = {Phys. Rev. B},
  volume = {83},
  issue = {24},
  pages = {245132},
  numpages = {39},
  year = {2011},
  month = {Jun},
  publisher = {American Physical Society},
  doi = {10.1103/PhysRevB.83.245132},
  url = {https://link.aps.org/doi/10.1103/PhysRevB.83.245132}
}

@article{PhysRevX.7.031057,
  title = {Exploring {Interacting Topological Insulators} with {Ultracold Atoms: The Synthetic Creutz-Hubbard Model}},
  author = {J\"unemann, J. and Piga, A. and Ran, S.-J. and Lewenstein, M. and Rizzi, M. and Bermudez, A.},
  journal = {Phys. Rev. X},
  volume = {7},
  issue = {3},
  pages = {031057},
  numpages = {25},
  year = {2017},
  month = {Sep},
  publisher = {American Physical Society},
  doi = {10.1103/PhysRevX.7.031057},
  url = {https://link.aps.org/doi/10.1103/PhysRevX.7.031057}
}

@article{10.1063/5.0150118,
    author = {Dabiri, S. Sajad and Cheraghchi, Hosein},
    title = {Electric circuit simulation of {Floquet} topological insulators in {Fourier} space},
    journal = {Journal of Applied Physics},
    volume = {134},
    number = {8},
    pages = {084303},
    year = {2023},
    month = {08},
    issn = {0021-8979},
    doi = {10.1063/5.0150118},
    url = {https://doi.org/10.1063/5.0150118},
}

@article{Grifoni1998,
  title = {Driven quantum tunnelling},
  author = {Grifoni, Milena and H\"{a}nggi, Peter},
  journal = {Phys. Rep.},
  volume = {304},
  pages = {229--354},
  year = {1998},
  publisher = {Elsevier},
  doi = {10.1016/S0370-1573(98)00022-2},
  url = {https://www.sciencedirect.com/science/article/pii/S0370157398000222}
}

@article{Goldman2014,
  title = {Periodically {Driven Quantum Systems: Effective Hamiltonians} and {Engineered Gauge Fields}},
  author = {Goldman, N. and Dalibard, J.},
  journal = {Phys. Rev. X},
  volume = {4},
  issue = {3},
  pages = {031027},
  numpages = {29},
  year = {2014},
  month = {Sep},
  publisher = {American Physical Society},
  doi = {10.1103/PhysRevX.4.031027},
  url = {https://link.aps.org/doi/10.1103/PhysRevX.4.031027}
}

@article{Restrepo2016,
  title = {{Driven Open Quantum Systems and Floquet Stroboscopic Dynamics}},
  author = {Restrepo, Santiago and Cerrillo, Javier and Bastidas, Victor M. and Angelakis, Dimitris G. and Brandes, Tobias},
  journal = {Phys. Rev. Lett.},
  volume = {117},
  issue = {25},
  pages = {250401},
  numpages = {6},
  year = {2016},
  month = {Dec},
  publisher = {American Physical Society},
  doi = {10.1103/PhysRevLett.117.250401},
  url = {https://link.aps.org/doi/10.1103/PhysRevLett.117.250401}
}

@article{Cayssol2013,
  title = {Floquet topological insulators},
  author = {Cayssol, J. and D\'{o}ra, B. and Simon, F. and Moessner, R.},
  journal = {Phys. Status Solidi RRL},
  volume = {7},
  issue = {1-2},
  pages = {101--108},
  year = {2013},
  month = {Jan},
  publisher = {Wiley},
  doi = {10.1002/pssr.201206451},
  url = {https://onlinelibrary.wiley.com/doi/abs/10.1002/pssr.201206451}
}

@article{Rudner2013,
  title = {{Anomalous Edge States and the Bulk-Edge Correspondence for Periodically Driven Two-Dimensional Systems}},
  author = {Rudner, M. S. and Lindner, N. H. and Berg, E. and Levin, M.},
  journal = {Phys. Rev. X},
  volume = {3},
  issue = {3},
  pages = {031005},
  numpages = {19},
  year = {2013},
  month = {Jul},
  publisher = {American Physical Society},
  doi = {10.1103/PhysRevX.3.031005},
  url = {https://link.aps.org/doi/10.1103/PhysRevX.3.031005}
}

@article{GomezLeon2013,
  title = {{Floquet-Bloch Theory and Topology in Periodically Driven Lattices}},
  author = {G\'{o}mez-Le\'{o}n, \'{A}lvaro and Platero, Gloria},
  journal = {Phys. Rev. Lett.},
  volume = {110},
  issue = {20},
  pages = {200403},
  numpages = {5},
  year = {2013},
  month = {May},
  publisher = {American Physical Society},
  doi = {10.1103/PhysRevLett.110.200403},
  url = {https://link.aps.org/doi/10.1103/PhysRevLett.110.200403}
}

@article{PhysRevB.108.L220301,
  title = {{Loss-induced Floquet non-Hermitian skin effect}},
  author = {Li, Yaohua and Lu, Cuicui and Zhang, Shuang and Liu, Yong-Chun},
  journal = {Phys. Rev. B},
  volume = {108},
  issue = {22},
  pages = {L220301},
  numpages = {6},
  year = {2023},
  month = {Dec},
  publisher = {American Physical Society},
  doi = {10.1103/PhysRevB.108.L220301},
  url = {https://link.aps.org/doi/10.1103/PhysRevB.108.L220301}
}

@article{Ji_2024,
doi = {10.1088/1361-648X/ad2c73},
url = {https://dx.doi.org/10.1088/1361-648X/ad2c73},
year = {2024},
month = {mar},
publisher = {IOP Publishing},
volume = {36},
number = {24},
pages = {243001},
author = {Ji, Xiang and Yang, Xiaosen},
title = {{Generalized bulk-boundary correspondence in periodically driven non-Hermitian systems}},
journal = {Journal of Physics: Condensed Matter}
}

@article{PhysRevLett.132.063804,
  title = {{Photonic Floquet Skin-Topological Effect}},
  author = {Sun, Yeyang and Hou, Xiangrui and Wan, Tuo and Wang, Fangyu and Zhu, Shiyao and Ruan, Zhichao and Yang, Zhaoju},
  journal = {Phys. Rev. Lett.},
  volume = {132},
  issue = {6},
  pages = {063804},
  numpages = {6},
  year = {2024},
  month = {Feb},
  publisher = {American Physical Society},
  doi = {10.1103/PhysRevLett.132.063804},
  url = {https://link.aps.org/doi/10.1103/PhysRevLett.132.063804}
}

@article{Rudner2020,
  title = {{Band structure engineering and non-equilibrium dynamics in Floquet topological insulators}},
  author = {Rudner, M. S. and Lindner, N. H.},
  journal = {Nat. Rev. Phys.},
  volume = {2},
  pages = {229--244},
  year = {2020},
  month = {Apr},
  publisher = {Springer Nature},
  doi = {10.1038/s42254-020-0170-z},
  url = {https://doi.org/10.1038/s42254-020-0170-z}
}

@article{Agrawal2022,
  title = {{Floquet topological phases with high Chern numbers in a periodically driven extended Su–Schrieffer–Heeger model}},
  author = {Agrawal, Anurag and Bandyopadhyay, Jayendra N.},
  journal = {J. Phys.: Condens. Matter},
  volume = {34},
  pages = {305401},
  year = {2022},
  publisher = {IOP Publishing},
  doi = {10.1088/1361-648X/ac6eac},
  url = {https://iopscience.iop.org/article/10.1088/1361-648X/ac6eac}
}

@article{Li2020,
  title = {{Topological Properties of an Extend Su–Schrieffer–Heeger Model Under Periodic Kickings}},
  author = {Li, Chun-Fang and Luan, Li-Ning and Wang, Li-Chen},
  journal = {Int. J. Theor. Phys.},
  volume = {59},
  pages = {2852--2861},
  year = {2020},
  publisher = {Springer},
  doi = {10.1007/s10773-020-04545-7},
  url = {https://link.springer.com/article/10.1007/s10773-020-04545-7}
}

@article{Roy2023,
  title = {{Topological properties of a periodically driven Creutz ladder}},
  author = {Roy, Koustav and Basu, Saurabh},
  journal = {Phys. Rev. B},
  volume = {108},
  pages = {045415},
  year = {2023},
  publisher = {American Physical Society},
  doi = {10.1103/PhysRevB.108.045415},
  url = {https://journals.aps.org/prb/abstract/10.1103/PhysRevB.108.045415}
}

@article{Yang2022,
  title = {{Observation of Floquet topological phases with large Chern numbers}},
  author = {Yang, Kai and Xu, Sheng and Zhou, Ling and Zhao, Ziran and Xie, Tongxin and Ding, Zhenyu and Ma, Wenlong and Gong, Jiangbin and Shi, Fusheng and Du, Jiangfeng},
  journal = {Phys. Rev. B},
  volume = {106},
  pages = {184106},
  year = {2022},
  publisher = {American Physical Society},
  doi = {10.1103/PhysRevB.106.184106},
  url = {https://journals.aps.org/prb/abstract/10.1103/PhysRevB.106.184106}
}

@article{Roy2024,
  title = {{Single and multifrequency driving protocols in a Rashba nanowire proximitized to an $s$-wave superconductor}},
  author = {Roy, Koustav and Basu, Saurabh},
  journal = {Phys. Rev. B},
  volume = {110},
  pages = {165403},
  year = {2024},
  publisher = {American Physical Society},
  doi = {10.1103/PhysRevB.110.165403},
  url = {https://arxiv.org/abs/2405.11307}
}

@article{SanJose2013prime,
  title = {{Multiple Andreev reflection and critical current in topological superconducting nanowire junctions}},
  author = {San-Jose, Pablo and Cayao, Jorge and Prada, Elsa and Aguado, Ramón},
  journal = {New J. Phys.},
  volume = {15},
  pages = {075019},
  year = {2013},
  publisher = {IOP Publishing},
  doi = {10.1088/1367-2630/15/7/075019},
  url = {https://iopscience.iop.org/article/10.1088/1367-2630/15/7/075019}
}

@article{Peng2021prime,
  title = {{Floquet Majorana bound states in voltage-biased planar Josephson junctions}},
  author = {Peng, Cheng and Haim, Ariel and Karzig, Torsten and Peng, Yujing and Refael, Gil},
  journal = {Phys. Rev. Research},
  volume = {3},
  pages = {023108},
  year = {2021},
  publisher = {American Physical Society},
  doi = {10.1103/PhysRevResearch.3.023108},
  url = {https://journals.aps.org/prresearch/abstract/10.1103/PhysRevResearch.3.023108}
}

@article{PhysRevLett.119.093901,
  title = {{Experimental Realization of Floquet $\mathcal{P}\mathcal{T}$-Symmetric Systems}},
  author = {Chitsazi, Mahboobeh and Li, Huanan and Ellis, F. M. and Kottos, Tsampikos},
  journal = {Phys. Rev. Lett.},
  volume = {119},
  issue = {9},
  pages = {093901},
  numpages = {6},
  year = {2017},
  month = {Sep},
  publisher = {American Physical Society},
  doi = {10.1103/PhysRevLett.119.093901},
  url = {https://link.aps.org/doi/10.1103/PhysRevLett.119.093901}
}

@article{Liu2019prime,
  title = {{Floquet Majorana zero and $\pi$ modes in planar Josephson junctions}},
  author = {Liu, D. T. and Shabani, J. and Mitra, A.},
  journal = {Phys. Rev. B},
  volume = {99},
  pages = {094303},
  year = {2019},
  publisher = {American Physical Society},
  doi = {10.1103/PhysRevB.99.094303},
  url = {https://journals.aps.org/prb/abstract/10.1103/PhysRevB.99.094303}
}

@article{Roy2025,
  title = {{Floquet-engineered diode performance in a Majorana-quantum dot Josephson junction}},
  author = {Roy, Koustav and Paul, Gourab and Debnath, Debika and Bhattacharyya, Kuntal and Basu, Saurabh},
  journal = {arXiv preprint},
  year = {2025},
  eprint = {2503.07428},
  archivePrefix = {arXiv},
  primaryClass = {cond-mat.mes-hall},
  url = {https://arxiv.org/abs/2503.07428}
}

@article{Roy2024prime,
  author = {Roy, K. and Roy, S. and Basu, S.},
  title = {{Quasiperiodic disorder induced critical phases in a periodically driven dimerized $p$-wave Kitaev chain}},
  journal = {Sci Rep},
  volume = {14},
  pages = {20603},
  year = {2024},
  doi = {10.1038/s41598-024-70995-2},
  url = {https://doi.org/10.1038/s41598-024-70995-2}
}

@article{Pan2020,
  author = {Pan, Y. and Wang, B.},
  title = {{Time-crystalline phases and period-doubling oscillations in one-dimensional Floquet topological insulators}},
  journal = {Phys. Rev. Res.},
  volume = {2},
  pages = {043239},
  year = {2020},
  doi = {10.1103/PhysRevResearch.2.043239},
  url = {https://link.aps.org/doi/10.1103/PhysRevResearch.2.043239}
}

@article{Wang2022,
  author = {Wang, B. and Quan, J. and Han, J. and Shen, X. and Wu, H. and Pan, Y.},
  title = {{Observation of Photonic Topological Floquet Time Crystals}},
  journal = {Laser Photon. Rev.},
  volume = {16},
  pages = {2100469},
  year = {2022},
  doi = {10.1002/lpor.202100469},
  url = {https://onlinelibrary.wiley.com/doi/abs/10.1002/lpor.202100469}
}

@article{Kunst2018,
  author = {Kunst, F. K. and Edvardsson, E. and Budich, J. C. and Bergholtz, E. J.},
  title = {{Biorthogonal Bulk-Boundary Correspondence in Non-Hermitian Systems}},
  journal = {Phys. Rev. Lett.},
  volume = {121},
  pages = {026808},
  year = {2018},
  doi = {10.1103/PhysRevLett.121.026808},
  url = {https://doi.org/10.1103/PhysRevLett.121.026808}
}

@article{Herviou2019,
  author = {Herviou, L. and Bardarson, J. H. and Regnault, N.},
  title = {{Defining a bulk-edge correspondence for non-Hermitian Hamiltonians via singular-value decomposition}},
  journal = {Phys. Rev. A},
  volume = {99},
  pages = {052118},
  year = {2019},
  doi = {10.1103/PhysRevA.99.052118},
  url = {https://doi.org/10.1103/PhysRevA.99.052118}
}

@Article{Xiao2020,
author={Xiao, Lei
and Deng, Tianshu
and Wang, Kunkun
and Zhu, Gaoyan
and Wang, Zhong
and Yi, Wei
and Xue, Peng},
title={{Non-Hermitian bulk--boundary correspondence in quantum dynamics}},
journal={Nature Physics},
year={2020},
month={Jul},
day={01},
volume={16},
number={7},
issn={1745-2481},
doi={10.1038/s41567-020-0836-6},
url={https://doi.org/10.1038/s41567-020-0836-6}
}

@article{PhysRevB.103.075126,
  title = {{Non-Hermitian bulk-boundary correspondence in a periodically driven system}},
  author = {Cao, Yang and Li, Yang and Yang, Xiaosen},
  journal = {Phys. Rev. B},
  volume = {103},
  issue = {7},
  pages = {075126},
  numpages = {10},
  year = {2021},
  month = {Feb},
  publisher = {American Physical Society},
  doi = {10.1103/PhysRevB.103.075126},
  url = {https://link.aps.org/doi/10.1103/PhysRevB.103.075126}
}

@article{Ge2019,
  author = {Ge, Z.-Y. and Zhang, Y.-R. and Liu, T. and Li, S.-W. and Fan, H. and Nori, F.},
  title = {{Topological band theory for non-Hermitian systems from the Dirac equation}},
  journal = {Phys. Rev. B},
  volume = {100},
  pages = {054105},
  year = {2019},
  doi = {10.1103/PhysRevB.100.054105},
  url = {https://doi.org/10.1103/PhysRevB.100.054105}
}

@article{Zhang2020,
  author = {Zhang, K. and Yang, Z. and Fang, C.},
  title = {{Correspondence between Winding Numbers and Skin Modes in Non-Hermitian Systems}},
  journal = {Phys. Rev. Lett.},
  volume = {125},
  pages = {126402},
  year = {2020},
  doi = {10.1103/PhysRevLett.125.126402},
  url = {https://doi.org/10.1103/PhysRevLett.125.126402}
}

@article{Wu2020,
  author = {Wu, H. and An, J.-H.},
  title = {{Floquet topological phases of non-Hermitian systems}},
  journal = {Phys. Rev. B},
  volume = {102},
  pages = {041119},
  year = {2020},
  doi = {10.1103/PhysRevB.102.041119},
  url = {https://doi.org/10.1103/PhysRevB.102.041119}
}

@article{Creutz1999,
  author = {Creutz, M.},
  title = {{End States, Ladder Compounds, and Domain-Wall Fermions}},
  journal = {Phys. Rev. Lett.},
  volume = {83},
  pages = {2636},
  year = {1999},
  doi = {10.1103/PhysRevLett.83.2636},
  url = {https://link.aps.org/doi/10.1103/PhysRevLett.83.2636}
}

@article{Gholizadeh2018,
  author = {Gholizadeh, S. and Yahyavi, M. and Hetényi, B.},
  title = {Extended {Creutz} ladder with spin-orbit coupling: A one-dimensional analog of the Kane-Mele model},
  journal = {Europhys. Lett.},
  volume = {122},
  pages = {27001},
  year = {2018},
  doi = {10.1209/0295-5075/122/27001},
  url = {https://iopscience.iop.org/article/10.1209/0295-5075/122/27001/meta}
}

@article{Kang2020,
  author = {Kang, J. H. and Han, J. H. and Shin, Y.},
  title = {Creutz ladder in a resonantly shaken 1{D} optical lattice},
  journal = {New J. Phys.},
  volume = {22},
  pages = {013023},
  year = {2020},
  doi = {10.1088/1367-2630/ab61d7},
  url = {https://iopscience.iop.org/article/10.1088/1367-2630/ab61d7}
}

@article{Li2013,
  author = {Li, X. and Zhao, E. and Liu, W. V.},
  title = {Topological states in a ladder-like optical lattice containing ultracold atoms in higher orbital bands},
  journal = {Nat. Commun.},
  volume = {4},
  pages = {1523},
  year = {2013},
  doi = {10.1038/ncomms2523},
  url = {https://www.nature.com/articles/ncomms2523}
}

@article{Hugel2014,
  author = {Hügel, D. and Paredes, B.},
  title = {Chiral ladders and the edges of quantum {Hall} insulators},
  journal = {Phys. Rev. A},
  volume = {89},
  pages = {023619},
  year = {2014},
  doi = {10.1103/PhysRevA.89.023619},
  url = {https://link.aps.org/doi/10.1103/PhysRevA.89.023619}
}

@article{Kuno2020,
  author = {Kuno, Y.},
  title = {{Extended flat band, entanglement, and topological properties in a Creutz ladder}},
  journal = {Phys. Rev. B},
  volume = {101},
  pages = {184112},
  year = {2020},
  doi = {10.1103/PhysRevB.101.184112},
  url = {https://journals.aps.org/prb/abstract/10.1103/PhysRevB.101.184112}
}

@article{DiLiberto2019,
  author = {Di Liberto, M. and Mukherjee, S. and Goldman, N.},
  title = {Nonlinear dynamics of {Aharonov-Bohm} cages},
  journal = {Phys. Rev. A},
  volume = {100},
  pages = {043829},
  year = {2019},
  doi = {10.1103/PhysRevA.100.043829},
  url = {https://journals.aps.org/pra/abstract/10.1103/PhysRevA.100.043829}
}

@article{Junemann2017,
  author = {Jünemann, J. and Piga, A. and Ran, S.-J. and Lewenstein, M. and Rizzi, M. and Bermudez, A.},
  title = {Exploring {Interacting} {Topological} {Insulators} with {Ultracold} {Atoms}: {The} {Synthetic} {Creutz-Hubbard} {Model}},
  journal = {Phys. Rev. X},
  volume = {7},
  pages = {031057},
  year = {2017},
  doi = {10.1103/PhysRevX.7.031057},
  url = {https://link.aps.org/doi/10.1103/PhysRevX.7.031057}
}

@article{LiangLi2022,
  author = {Liang, H.-Q. and Li, L.},
  title = {{Topological properties of non-Hermitian Creutz ladders}},
  journal = {Chinese Phys. B},
  volume = {31},
  pages = {010310},
  year = {2022},
  doi = {10.1088/1674-1056/ac3991},
  url = {https://iopscience.iop.org/article/10.1088/1674-1056/ac3991}
}

@article{Zhou2020,
  author = {Zhou, L.},
  title = {{Non-Hermitian Floquet Phases with Even-Integer Topological Invariants in a Periodically Quenched Two-Leg Ladder}},
  journal = {Entropy},
  volume = {22},
  number = {7},
  pages = {746},
  year = {2020},
  doi = {10.3390/e22070746},
  url = {https://doi.org/10.3390/e22070746}
}

@article{SuSchriefferHeeger1979,
  author = {Su, W. P. and Schrieffer, J. R. and Heeger, A. J.},
  title = {Solitons in {Polyacetylene}},
  journal = {Phys. Rev. Lett.},
  volume = {42},
  pages = {1698--1701},
  year = {1979},
  doi = {10.1103/PhysRevLett.42.1698},
  url = {https://doi.org/10.1103/PhysRevLett.42.1698}
}

@article{Roy2025Creutz,
  title = {{Topological characterization of a non-Hermitian ladder via Floquet non-Bloch theory}},
  author = {Roy, Koustav and Gogoi, Koustabh and Basu, Saurabh},
  journal = {Phys. Rev. B},
  volume = {111},
  issue = {11},
  pages = {115424},
  numpages = {18},
  year = {2025},
  month = {Mar},
  publisher = {American Physical Society},
  doi = {10.1103/PhysRevB.111.115424},
  url = {https://link.aps.org/doi/10.1103/PhysRevB.111.115424}
}

@article{EckardtAnisimovas2015,
  author = {Eckardt, A. and Anisimovas, E.},
  title = {{High-frequency approximation for periodically driven quantum systems from a Floquet-space perspective}},
  journal = {New J. Phys.},
  volume = {17},
  pages = {093039},
  year = {2015},
  doi = {10.1088/1367-2630/17/9/093039},
  url = {https://doi.org/10.1088/1367-2630/17/9/093039}
}

@article{Benito2014,
  author = {Benito, M. and Gómez-León, A. and Bastidas, V. M. and Brandes, T. and Platero, G.},
  title = {{Floquet engineering of long-range $p$-wave superconductivity}},
  journal = {Phys. Rev. B},
  volume = {90},
  pages = {205127},
  year = {2014},
  doi = {10.1103/PhysRevB.90.205127},
  url = {https://doi.org/10.1103/PhysRevB.90.205127}
}

@article{Wang2020,
  author = {Wang, H.-Y. and Zhuang, L. and Gao, X.-L. and Zhao, X.-D. and Liu, W.-M.},
  title = {{Robust Majorana edge modes with low frequency multiple time periodic driving}},
  journal = {J. Phys.: Condens. Matter},
  volume = {32},
  pages = {355404},
  year = {2020},
  doi = {10.1088/1361-648X/ab8ddd},
  url = {https://doi.org/10.1088/1361-648X/ab8ddd}
}

@article{Lin2023,
  author = {Lin, R. and Tai, T. and Li, L. and others},
  title = {Topological non-Hermitian skin effect},
  journal = {Front. Phys.},
  volume = {18},
  pages = {53605},
  year = {2023},
  doi = {10.1007/s11467-023-1309-z},
  url = {https://doi.org/10.1007/s11467-023-1309-z}
}

@article{YokomizoMurakami2021,
  author = {Yokomizo, K. and Murakami, S.},
  title = {Non-Bloch band theory in bosonic Bogoliubov–de Gennes systems},
  journal = {Phys. Rev. B},
  volume = {103},
  pages = {165123},
  year = {2021},
  doi = {10.1103/PhysRevB.103.165123},
  url = {https://doi.org/10.1103/PhysRevB.103.165123}
}

@article{Asboth2014,
  author = {Asbóth, J. K. and Tarasinski, B. and Delplace, P.},
  title = {{Chiral symmetry and bulk-boundary correspondence in periodically driven one-dimensional systems}},
  journal = {Phys. Rev. B},
  volume = {90},
  pages = {125143},
  year = {2014},
  doi = {10.1103/PhysRevB.90.125143},
  url = {https://link.aps.org/doi/10.1103/PhysRevB.90.125143}
}

@article{AsbothObuse2013,
  author = {Asbóth, J. K. and Obuse, H.},
  title = {{Bulk-boundary correspondence for chiral symmetric quantum walks}},
  journal = {Phys. Rev. B},
  volume = {88},
  pages = {121406(R)},
  year = {2013},
  doi = {10.1103/PhysRevB.88.121406},
  url = {https://link.aps.org/doi/10.1103/PhysRevB.88.121406}
}

@article{RoyHarper2017,
  author = {Roy, R. and Harper, F.},
  title = {{Periodic table for Floquet topological insulators}},
  journal = {Phys. Rev. B},
  volume = {96},
  pages = {155118},
  year = {2017},
  doi = {10.1103/PhysRevB.96.155118},
  url = {https://link.aps.org/pdf/10.1103/PhysRevB.96.155118}
}

@misc{footnote,
  howpublished = {The task of finding GBZ is best accomplished by writing the Hamiltonian in the form of a massless Dirac equation, namely, $H(k)=\vec{d}(k)\cdot\vec{\sigma}$, where $\vec{\sigma}$ denotes the Pauli matrices}
}

@article{Ji2025,
  title = {{Floquet engineering of point‐gapped topological phases in classical and quantum systems}},  
  author = {Ji, X. and others},  
  journal = {Physical Review B},  
  volume = {111},  
  number = {19},  
  pages = {195419},  
  year = {2025},  
  month = {May},
  doi = {10.1103/PhysRevB.111.195419},
  url = {https://doi.org/10.1103/PhysRevB.111.195419}
}

@article{Manna2023,
  title = {{Inner skin effects on non-Hermitian topological fractals}},
  author = {Manna, S. and Roy, B.},
  journal = {Communications Physics},
  volume = {6},
  number = {1},
  pages = {10},
  year = {2023},
  publisher = {Nature Publishing Group},
  doi = {10.1038/s42005-023-01130-2},
  url = {https://doi.org/10.1038/s42005-023-01130-2}
}

@article{Ryu2010,
  title={Topological insulators and superconductors: tenfold way and dimensional hierarchy},
  author={Ryu, Shinsei and Schnyder, Andreas P. and Furusaki, Akira and Ludwig, Andreas W. W.},
  journal={New Journal of Physics},
  volume={12},
  number={6},
  pages={065010},
  year={2010},
  publisher={IOP Publishing},
  doi={10.1088/1367-2630/12/6/065010},
  url={https://iopscience.iop.org/article/10.1088/1367-2630/12/6/065010/meta}
}

@article{Schnyder2008,
  title={Classification of topological insulators and superconductors in three spatial dimensions},
  author={Schnyder, Andreas P. and Ryu, Shinsei and Furusaki, Akira and Ludwig, Andreas W. W.},
  journal={Physical Review B},
  volume={78},
  number={19},
  pages={195125},
  year={2008},
  publisher={American Physical Society},
  doi={10.1103/PhysRevB.78.195125},
  url={https://link.aps.org/doi/10.1103/PhysRevB.78.195125}
}

@article{PhysRevApplied.18.054034,
  title = {On-Demand Parity-Time Symmetry in a Lone Oscillator through Complex Synthetic Gauge Fields},
  author = {Quiroz-Ju\'arez, Mario A. and Agarwal, Kaustubh S. and Cochran, Zachary A. and Arag\'on, Jos\'e L. and Joglekar, Yogesh N. and Le\'on-Montiel, Roberto de J.},
  journal = {Phys. Rev. Appl.},
  volume = {18},
  issue = {5},
  pages = {054034},
  numpages = {9},
  year = {2022},
  month = {Nov},
  publisher = {American Physical Society},
  doi = {10.1103/PhysRevApplied.18.054034},
  url = {https://link.aps.org/doi/10.1103/PhysRevApplied.18.054034}
}

@misc{supp,
  title = {See {S}upplemental {M}aterial at [URL], which includes Refs. [21] and [83].},
}
\bibliographystyle{apsrev4-2}

\end{document}